\pdfoutput=1
\documentclass[preprint,12pt,authoryear]{elsarticle}

\usepackage[T1]{fontenc}
\usepackage[utf8]{inputenc}

\usepackage{amsmath,amssymb,amsthm,mathtools}
\usepackage{bm}
\usepackage{enumitem}
\usepackage{booktabs}
\usepackage{multirow}
\usepackage{array}
\usepackage{adjustbox}
\usepackage{graphicx}
\usepackage{caption}
\usepackage[table]{xcolor}
\usepackage{hyperref}
\usepackage{arydshln}
\usepackage{etoc}

\renewcommand{\bottomfraction}{0.9}
\renewcommand{\textfraction}{0.1}

\hypersetup{
  colorlinks = true,
  citecolor  = violet,
  linkcolor  = red,
  urlcolor   = blue
}

\makeatletter
\def\ps@pprintTitle{\let\@oddhead\@empty\let\@evenhead\@empty\let\@oddfoot\@empty\let\@evenfoot\@oddfoot}
\makeatother

\newtheorem{theorem}{Theorem}
\newtheorem{proposition}{Proposition}
\newtheorem{lemma}{Lemma}
\newtheorem{definition}{Definition}
\newtheorem{remark}{Remark}
\newtheorem{assumption}{Assumption}

\newcommand{\R}{\mathbb{R}}
\newcommand{\E}{\mathbb{E}}
\newcommand{\Var}{\mathrm{Var}}
\newcommand{\Cov}{\mathrm{Cov}}
\newcommand{\sd}{\mathrm{sd}}
\newcommand{\one}{\mathbf{1}}
\newcommand{\SemiVar}{\mathrm{S}}
\newcommand{\SemiVol}{\mathrm{S}}
\newcommand{\SemiVolP}{\mathrm{S}_{P}}
\newcommand{\CVaR}{\mathrm{CVaR}}
\newcommand{\DR}{\mathrm{DR}}
\newcommand{\Sortino}{\mathrm{Sortino}}
\newcommand{\Sharpe}{\mathrm{Sharpe}}
\newcommand{\Rachev}{\mathrm{Rachev}}
\newcommand{\MaxDD}{\mathrm{MaxDD}}
\newcommand{\Calmar}{\mathrm{Calmar}}
\newcommand{\UI}{\mathrm{UI}}
\newcommand{\Turn}{\mathrm{Turn}}
\newcommand{\RC}{\mathrm{RC}}
\newcommand{\dd}{\mathrm{dd}}

\begin{document}
\etocdepthtag.toc{mtmain}

\begin{frontmatter}

\title{Semivariance and Semivolatility in Portfolio Selection:
A Unified Framework for Downside-Risk Portfolio Models}

\author[uni]{Francesco Cesarone\corref{cor1}}
\ead{francesco.cesarone@uniroma3.it}

\author[uni]{Massimiliano Corradini}
\ead{massimiliano.corradini@uniroma3.it}

\author[uni]{Nicol\`o Giunta}
\ead{nicolo.giunta@uniroma3.it}

\cortext[cor1]{Corresponding author.}

\affiliation[uni]{
organization={Roma Tre University, Department of Business Economics}, addressline={Via Silvio D'Amico, 77}, city={Rome}, postcode={00145}, country={Italy} }
\begin{abstract}
In this paper, we study downside risk in portfolio selection beyond minimum-risk models, where semivariance and its square root, semivolatility, yield the same portfolio.
Downside-risk models are often presented as direct extensions of their variance-based counterparts, although different classes of portfolio problems require different properties of the risk measure.
We classify these models by the degree of positive homogeneity of the risk measure in the weights: degree one makes risk contributions add up to the portfolio risk and keeps diversification and reward-to-risk ratios scale invariant.
For three thresholds, the portfolio mean, zero and a market-index return, we state a non-degeneracy condition on long-only downside risk under which the risk-parity, most-diversified and Sortino-ratio problems are well posed.
Under this condition, the two ratio problems reduce to convex programs, and the differentiability of semivolatility ensures that the risk-budgeting formulation attains the prescribed budgets.
We also characterize when below-mean semivolatility is additive.
Finally, we provide an out-of-sample analysis on two real-world datasets of total returns, the current constituents of the Dow Jones and Nasdaq-100 indexes.
We compare seventeen long-only strategies, covering the minimum-risk, reward-to-risk, risk-parity and most-diversified families, in a rolling-window backtest, and test the differences within each family by block-bootstrap inference.
\end{abstract}


\end{frontmatter}

\section{Introduction}

Since the seminal contribution of \citet{markowitz1952}, portfolio selection has developed mainly within the mean--variance paradigm, whose convex quadratic programs are tractable; variance thus remains the standard benchmark in most portfolio selection models \citep{kolm2014sixty}.

Variance, however, penalizes upward and downward deviations symmetrically, while investors are typically more concerned with unfavorable outcomes.
Hence the interest in downside-risk measures, of which semivariance, accounting only for negative deviations, is the most widely discussed alternative \citep{markowitz1959,hogan1974toward,bawa1975optimal,bawa1976admissible,fishburn1977mean,nawrocki1999brief,estrada2006downside,cesarone2025semivariance}.
The other main strand of non-variance risk measures is the linear-programming one, initiated by the Mean Absolute Deviation (MAD) model of \citet{konno1991mean}; see \citet{mansini2014twenty}.

Semivariance has traditionally been considered less convenient than variance \citep{estrada2008mean,jin2006note,yao2021general}, although on scenario data mean--semivariance problems have long been solvable by quadratic programming \citep{markowitz1993computation}, and recent results show that semivariance, semicovariance matrices and lower partial moment models admit a more structured treatment \citep{cesarone2025semivariance}.
Downside-risk models are often presented as immediate extensions of their variance-based counterparts, without making explicit that different classes of portfolio problems require different structural properties of the risk measure, with concrete consequences outside the minimum-risk case.

Risk budgeting is well established for volatility \citep{maillard2010properties,bruder2012managing,spinu2013,roncalli2013introduction}, has been extended to CVaR \citep{boudt2013asset,cesarone2017minimum}, and \citet{bruder2012managing} observe, without proof, that its logarithmic formulation extends to any convex risk measure admitting an Euler decomposition.
Outside long-only volatility, tractability has been established one formulation at a time: risk parity is cast as a \emph{non-convex} least-squares problem \citep{bai2016least}, attacked by evolutionary heuristics once the long-only structure is dropped \citep{hochreiter2015evolutionary}, and settled only recently \citep{cetingoz2024risk}.
Diversification-ratio maximization becomes non-convex and calls for global optimization machinery \citep{barkhagen2023optimising}, while downside performance ratios are reformulated individually, as for the Omega ratio of \citet{kapsos2014optimizing}.
The recurrent obstacle is thus the generality in which the models are posed, not the downside nature of the risk measure.

Downside-risk portfolio models should distinguish between \emph{degree-two} and \emph{degree-one} downside measures.
Semivariance is positively homogeneous of degree two in the weights, hence the natural downside counterpart of variance in minimum-risk formulations, where it shares its minimizer with its square root.
Semivolatility is positively homogeneous of degree one, and is the measure required by diversification ratios and performance ratios that are invariant to rescaling the weights.
Risk budgeting supports both degrees, but only degree one normalizes the risk contributions so that they add up to the risk.
The properties of below-mean semivolatility behind this distinction are those of the standard semideviation of \citet{ogryczak1999from,ogryczak2001consistency} (Section~\ref{subsec:semi}), and only their systematic use as an organizing principle is new.

The contributions of this paper are the following.
First, we make the threshold explicit in this degree-based taxonomy.
We give a non-degeneracy condition on long-only downside risk relative to a threshold, possibly random, such as a market-index return (Assumption~\ref{ass:nondeg-tau}), under which the fixed-target and benchmark-relative risk-parity, most-diversified and Sortino-ratio problems are well posed.
This condition does not follow from its below-mean analogue, which we characterize in Proposition~\ref{prop:nondeg-equiv}.
For the ratio models, convexity, positivity and degree-one homogeneity in the weights suffice, the threshold entering only through positivity.
They are what is required by the reduction of reward--risk ratios to convex programs \citep{schaible1974parameter,stoyanov2007optimal}, also applied to diversification ratios by \citet{cesarone2025return}, which we restate as Lemma~\ref{lem:schaible}.
As a special case of \citet{fischer2003risk}, portfolio semivolatility is continuously differentiable with an explicit gradient and, unlike its $L^1$ analogue, requires no continuity assumption on the return distribution (Lemma~\ref{lem:sv-differentiable}), so that the logarithmic program realizes the prescribed downside risk budgets (Proposition~\ref{prop:rp-budget}).
Second, we characterize the additivity of below-mean semivolatility (Lemma~\ref{lem:sv-additivity-pair} and Theorem~\ref{thm:sv-additivity-multi}), the semivolatility counterparts of \citet[Lemma~2 and Theorem~6]{ararat2021mad}, whose novel ingredient is the downside proportionality condition forced by the $L^2$ structure.
The closed-form risk-parity portfolio it yields (Proposition~\ref{prop:svrp-closed-form}) holds only for a universe with no downside diversification, a limiting case.
Section~S.7.1 of the Supplementary Material lists each theoretical result with its hypotheses, closest antecedent and status (recalled, specialized, extended or new).
Third, extending earlier comparisons of mean--variance and mean--downside-risk portfolios \citep{harlow1991asset,grootveld1999variance}, we compare seventeen long-only strategies out of sample on two real-world datasets of total returns over 18.6 and 14.1 years, respectively, and we test the differences within each family with block-bootstrap inference.
The design follows the comparison of minimum-risk and diversification strategies of \citet{cesarone2017minimum} and adds two axes to its model classes, the risk measure and the threshold.
Replacing volatility with semivolatility produces no significant difference in the Sharpe ratio, whereas the market-index threshold changes it by mostly larger but often imprecise amounts.

Section~\ref{sec:framework} introduces the theoretical framework, Section~\ref{sec:models} the portfolio models and the role of convexity and homogeneity in their tractability, Section~\ref{sec:empirical} the out-of-sample analysis, and Section~\ref{sec:conclusion} the conclusions.
\ref{app:additivity} collects the additivity results and~\ref{app:proofs} the proofs, while the Supplementary Material adds further tests and analyses.

\section{Theoretical framework}\label{sec:framework}

\subsection{Preliminary concepts}\label{subsec:setting}

Let \((\Omega,\mathcal{F},\mathbb{P})\) be a probability space and \(L^2 := L^2(\Omega,\mathcal{F},\mathbb{P})\) the space of square-integrable random variables, with positive cone $L^2_+:=\{Z\in L^2 : Z\ge 0\ \mathbb{P}\text{-a.s.}\}$.
For $z\in\R$, $z_+:=\max\{z,0\}$ and $z_-:=\max\{-z,0\}$, acting pointwise in $\omega$ on a random variable.

For $n$ assets with linear returns \(R_1,\dots,R_n \in L^2\) and expected returns \(\mu_i := \E[R_i]\), we set $R:=(R_1,\dots,R_n)^\top$ and $\mu:=(\mu_1,\dots,\mu_n)^\top$, and denote by $e_i$ the $i$-th canonical basis vector of $\R^n$.
For a portfolio \(x \in \R^n\), the linear return is
\[
R_P(x) := \sum_{i=1}^n x_i R_i = R^\top x,
\]
and a fully invested long-only portfolio lies in the simplex
\[
\Delta := \left\{x \in \R_+^n : \one^\top x = 1\right\}.
\]

The results hold for any $\mathbb{P}$, and in Section~\ref{sec:empirical} $\mathbb{P}$ is the empirical distribution of $D_L$ equally likely historical scenarios (Section~\ref{subsec:data}; look-back approach, see, e.g., \citealt{demiguel2009optimal,cesarone2025semivariance}).

Since $x\mapsto R_P(x)$ is linear, convexity and positive homogeneity of a risk functional $\rho:L^2 \to \R_+$ carry over to $x \mapsto \rho(R_P(x))$.\footnote{\emph{Risk measure} is meant in the broad sense, not in the axiomatic coherent one \citep{artzner1999coherent,rockafellar2006generalized}.}
A functional $f$ is \emph{positively homogeneous of degree $k>0$} if $f(\lambda x)=\lambda^k f(x)$ for every $\lambda\ge0$; for $k=1$ subadditivity and convexity are equivalent.

\subsection{Variance, volatility, semivariance and semivolatility}
\label{subsec:semi}

For $X\in L^2$, $\Var(X):=\E[(X-\E[X])^2]$ and $\sd(X):=\sqrt{\Var(X)}$; at the portfolio level $\Var(R_P(x))=x^\top\Sigma x$ and $\sigma_P(x):=\sqrt{x^\top\Sigma x}$, with $\Sigma:=\Cov(R)$ the covariance matrix of asset returns.
Standard deviation is a degree-one deviation measure in the sense of \citet{rockafellar2006generalized}, whereas variance is of degree two and not subadditive (Table~\ref{tab:properties}).
On semivariance, semicovariance matrices and lower partial moments, see \citet{cesarone2025semivariance}.

\begin{definition}[Semivariance]\label{def:semivariance}
For \(X \in L^2\) and a target \(\tau \in L^2\), the below-target semivariance and the below-mean semivariance are defined, respectively, by
\[
\SemiVar_\tau^2(X):=\E[(X-\tau)_-^2], \qquad \SemiVar_\mu^2(X):=\E[(X-\E[X])_-^2].
\]
At the portfolio level, \(\SemiVar_\mu^2(R_P(x))=\E[(R_P(x)-\mu^\top x)_-^2]\).
\end{definition}

A random target covers a deterministic threshold \(\tau\in\R\) \citep{bawa1975optimal,fishburn1977mean} and a stochastic benchmark such as the market-index return $R^I$ \citep{meade2011detection}, written $R^I_t$ in Section~\ref{sec:empirical}.
Both semivariances are convex in \(x\), a random target included, since \(u\mapsto u_-^2\) is convex and \(x\mapsto R_P(x)-\tau\) and \(x\mapsto R_P(x)-\mu^\top x\) are affine.

\begin{remark}\label{rem:nosemicov}
Unlike variance, below-mean semivariance admits in general no representation \(x^\top \Sigma^- x\) with a matrix \(\Sigma^-\) independent of the weights, since the negative part in \(\E[(R_P(x)-\mu^\top x)_-^2]\) acts on the portfolio return and not on the individual asset returns.
The semicovariance matrix is therefore endogenous, and an exact representation requires a matrix \(\Sigma^-(x)\) depending on the portfolio \citep{estrada2008mean,cheremushkin2009d,cesarone2025semivariance}.
\end{remark}

\begin{definition}[Semivolatility]\label{def:semivolatility}
For \(X \in L^2\) and a target \(\tau\in L^2\), the below-mean semivolatility and the below-target semideviation are defined, respectively, by
\[
\SemiVol_\mu(X):=\left(\E[(X-\E[X])_-^2]\right)^{1/2}, \qquad \SemiVol_\tau(X):=\left(\E[(X-\tau)_-^2]\right)^{1/2}.
\]
At the portfolio level, we write \(\SemiVolP(x):=\SemiVol_\mu(R_P(x))\).
\end{definition}

In the text, semivolatility refers to the below-mean case and semideviation to a generic target, while the label SVol (semivolatility) in the strategy names of Table~\ref{tab:strategies} covers both.
We recall that below-mean semivolatility is the \emph{standard semideviation} of \citet{ogryczak1999from,ogryczak2001consistency} and, at \(p=2\), the deviation component of the one-sided-moment coherent risk measures \(\rho_{p,a}(X)=-\E[X]+a\,\|(X-\E[X])_-\|_{L^p}\), \(a\in[0,1]\), of \citet{fischer2003risk}.

\begin{proposition}[Properties of semivolatility]
\label{prop:semivolatility-properties}
The mapping \(X \mapsto \SemiVol_\mu(X)\) is convex, positively homogeneous of degree \(1\), subadditive, and shift invariant.
\end{proposition}

\begin{proof}
\(\SemiVol_\mu(X)=\|(X-\E[X])_-\|_{L^2}\) composes the linear, shift-invariant map \(X\mapsto X-\E[X]\) with \(Z\mapsto\|Z_-\|_{L^2}\), which is positively homogeneous of degree one and convex, by pointwise convexity of \(z\mapsto z_-\), monotonicity of the norm on \(L^2_+\) and the triangle inequality; subadditivity then follows.
\end{proof}

As summarized in Table~\ref{tab:properties}, semivariance suits the minimum-risk formulations of Section~\ref{sec:models} and semivolatility the diversification and performance ratios, which are invariant to rescaling the weights only when the risk functional has degree one.

If \(R_P(x)-\mu^\top x\) is symmetric about zero for every \(x\in\R^n_+\), in particular if \(R\) is elliptically distributed \citep{fang1990symmetric}, then \(\SemiVol_\mu(R_P(x)) = \sd(R_P(x))/\sqrt2\) and the below-mean models of Section~\ref{sec:models} coincide with their variance-based counterparts.
The threshold-based specifications (\(\tau=0\), \(\tau_t=R^I_t\)) need not coincide, since their semideviation also depends on where the threshold lies relative to the mean.

\begin{table}[htbp]
\centering
\caption{Structural comparison of the risk measures considered in the paper.
\((\checkmark)\): risk contributions add up to twice the measure; the
risk-budgeting portfolio coincides with that of its square root.}
\label{tab:properties}
\small
\renewcommand{\arraystretch}{1.15}
\makebox[\textwidth][c]{%
\begin{adjustbox}{width=1.45\textwidth}
\begin{tabular}{>{\raggedright\arraybackslash}p{8.5cm}cccc}
\toprule
\textbf{Property}
& \textbf{Variance}
& \textbf{Volatility}
& \textbf{Below-mean semivariance}
& \textbf{Below-mean semivolatility} \\
\midrule
Nonnegativity                  & \checkmark & \checkmark & \checkmark & \checkmark \\
Positive homogeneity (degree)  & \(2\)      & \(1\)      & \(2\)      & \(1\)      \\
Convexity                      & \checkmark & \checkmark & \checkmark & \checkmark \\
Subadditivity                  & \(\times\) & \checkmark & \(\times\) & \checkmark \\
Shift invariance               & \checkmark & \checkmark & \checkmark & \checkmark \\
\addlinespace
\multicolumn{5}{l}{\textit{Portfolio applications}} \\
\cmidrule(lr){1-5}
Suitable for minimum-risk models                   & \checkmark & \checkmark & \checkmark & \checkmark \\
Suitable for risk-budgeting models                 & \((\checkmark)\) & \checkmark & \((\checkmark)\) & \checkmark \\
Suitable for diversification- and performance-ratio models & \(\times\) & \checkmark & \(\times\) & \checkmark \\
\bottomrule
\end{tabular}
\end{adjustbox}%
}
\end{table}

\subsection{Non-degeneracy of downside risk}\label{subsec:nondeg}

\begin{assumption}[Non-degeneracy of downside risk]\label{ass:nondeg}
The only vector $x \in \R^n_+$ such that
\[
\bigl(R_P(x) - \mu^\top x\bigr)_- = 0 \qquad \mathbb{P}\text{-a.s.}
\]
is $x = 0$ (equivalently, $\mathbb{P}\{R_P(x)<\mu^\top x\}>0$ for every $x\in\Delta$).
\end{assumption}

\begin{proposition}[Equivalent formulations of non-degeneracy]
\label{prop:nondeg-equiv}
The following statements are equivalent:
\begin{enumerate}[label=(\roman*)]
    \item Assumption~\ref{ass:nondeg} holds;
    \item the only $x \in \R^n_+$ such that $(R-\mu)^\top x = 0$
    $\mathbb{P}$-a.s.\ is $x = 0$;
    \item $\SemiVol_\mu\!\bigl(R_P(x)\bigr) > 0$ for every
    $x \in \R^n_+\setminus\{0\}$;
    \item $\sd\!\bigl(R_P(x)\bigr) > 0$ for every $x \in \R^n_+\setminus\{0\}$.
\end{enumerate}
\end{proposition}

\begin{proof}
See~\ref{app:proofs}.
\end{proof}

Formulation~(ii) is the non-degeneracy assumption of \citet{ararat2021mad}, and by~(iv) it holds if and only if no nonzero $x\ge0$ lies in the kernel of $\Sigma$, in particular whenever $\Sigma$ is positive definite.
Taking $x=e_i$ in~(iii) gives $\SemiVol_\mu(R_i)>0$ for every $i$, so that inverse semivolatility weights are well defined and almost surely constant assets, such as a risk-free one, are excluded.

For a target $\tau\in L^2$, on the simplex $R_P(x)-\tau=(R-\tau\one)^\top x$, so that $\SemiVol_\tau(R_P(x))=\rho_\tau(x)$ for $x\in\Delta$, where
\begin{equation}\label{eq:rho-tau}
\rho_\tau(x):=\bigl\|\bigl((R-\tau\one)^\top x\bigr)_-\bigr\|_{L^2},
\qquad x\in\R^n .
\end{equation}
The function $\rho_\tau$ is continuous, convex and positively homogeneous of degree one in the weights, even though $\SemiVol_\tau$ is not positively homogeneous in the portfolio return when $\tau\not\equiv0$.
Its positivity is a separate hypothesis.

\begin{assumption}[Non-degeneracy with respect to a target]\label{ass:nondeg-tau}
Let $\tau\in L^2$.
The only vector $x\in\R^n_+$ such that $\bigl((R-\tau\one)^\top x\bigr)_-=0$ $\mathbb{P}$-a.s.\ is $x=0$ (equivalently, $\mathbb{P}\{R_P(x)<\tau\}>0$ for every $x\in\Delta$).
\end{assumption}

Assumption~\ref{ass:nondeg-tau} does not follow from Assumption~\ref{ass:nondeg}: it fails if some non-constant $R_i\ge\tau$ a.s.\ or, for $\tau=R^I$, if some long-only portfolio never underperforms the index.
This happens, in particular, when the benchmark is exactly replicable, $R^I=R^\top\bar x$ for some $\bar x\in\Delta$, since then $\rho_{R^I}(\bar x)=0$.
Whenever Assumption~\ref{ass:nondeg-tau} fails, $\rho_\tau$ vanishes on a nonzero convex cone of long-only portfolios, and the threshold-based models of Section~\ref{sec:models} are ill posed, since the Sortino and diversification ratios can become unbounded and the risk-budgeting program degenerates.

\begin{remark}[Additivity of semivolatility]\label{rem:additivity}
In~\ref{app:additivity} we show that $\SemiVol_\mu(X+Y)=\SemiVol_\mu(X)+\SemiVol_\mu(Y)$ if and only if the downside deviations $(X-\E[X])_-$ and $(Y-\E[Y])_-$ are positively collinear (Lemma~\ref{lem:sv-additivity-pair}), with a multivariate extension and a common-downside-factor form (Theorem~\ref{thm:sv-additivity-multi}).
Since deviations are centered, collinearity implies that $X-\E[X]$ and $Y-\E[Y]$ never have opposite signs, the Mean Absolute Deviation additivity condition of \citet[Lemma~2]{ararat2021mad}.
Additivity of below-mean semivolatility is therefore strictly stronger than MAD additivity, as the examples in~\ref{app:additivity} show, and strictly weaker than volatility additivity, which requires positive collinearity of the full deviations.
\end{remark}

\section{Portfolio models under downside risk}\label{sec:models}

\subsection{Mean--variance versus mean--semivariance}\label{subsec:msv}

The classical mean--variance model with target return \(\eta\) is the convex quadratic program
\begin{equation}\label{prob:mv}
\begin{aligned}
\min_{x \in \R^n}\quad & x^\top \Sigma x \\
\text{s.t.}\quad
& \mu^\top x = \eta,\\
& \one^\top x = 1,\\
& x \ge 0.
\end{aligned}
\end{equation}
Its downside-risk counterpart is the mean--semivariance problem
\begin{equation}\label{prob:msv}
\begin{aligned}
\min_{x \in \R^n}\quad & \SemiVar_\mu^2(R_P(x)) \\
\text{s.t.}\quad
& \mu^\top x = \eta,\\
& \one^\top x = 1,\\
& x \ge 0,
\end{aligned}
\end{equation}
which is convex, though its objective is not a quadratic form with a fixed matrix (Remark~\ref{rem:nosemicov}).
Solution methods and analytical properties are given in \citet{markowitz1959,jin2006note,cesarone2025semivariance}.
The target-based variant, with $\SemiVar_\tau^2$ in place of $\SemiVar_\mu^2$, remains convex even for a random $\tau$ and gives the strategies MSV$_0$ and MSV$_{R^I}$.
In Section~\ref{sec:empirical} these models and the most-diversified portfolios of Section~\ref{subsec:mdp} are solved without the expected-return constraint.

\subsection{Risk parity under semivolatility}\label{subsec:rp}

Risk contributions derive from the Euler decomposition of positively homogeneous functionals \citep{maillard2010properties,roncalli2013introduction}, by which a differentiable $f$ of degree $k$ satisfies $\sum_i x_i\,\partial_i f(x)=k\,f(x)$, so that the contributions add up to the risk only for $k=1$, the case of semivolatility.

Let \(b_1,\dots,b_n>0\) be risk budgets with \(\sum_{i=1}^n b_i=1\).
Following the logarithmic formulation of \citet{maillard2010properties} (see also \citealt{bruder2012managing,spinu2013}), standard for degree-one risk budgeting \citep{roncalli2013introduction,cesarone2017minimum,bellini2021risk,ararat2021mad}, the semivolatility risk-budgeting problem reads
\begin{equation}\label{prob:rp-semivol}
\begin{aligned}
\min_{y \in \R^n}\quad & \SemiVolP(y) \\
\text{s.t.}\quad
& \sum_{i=1}^n b_i \log(y_i) \ge \gamma,\\
& y_i > 0, \qquad i=1,\dots,n,
\end{aligned}
\end{equation}
where \(\gamma \in \R\) fixes the scale of the solution.
The logarithmic constraint rules out \(y=0\) while keeping the feasible set convex, being the superlevel set of the concave function \(y\mapsto\sum_i b_i\log(y_i)\), so that Problem~\eqref{prob:rp-semivol} is a convex program on \(\R_{++}^n\) by Proposition~\ref{prop:semivolatility-properties}.

\citet[Sec.~2.4]{bruder2012managing} observe that this construction extends from volatility to any convex risk measure satisfying the Euler decomposition.
The construction presupposes differentiability, which convexity and homogeneity do not grant.
For the $L^1$ analogue of $\SemiVolP$, differentiability fails on discrete distributions at every portfolio whose return equals its mean in some scenario, as it does for CVaR \citep{mausser2018long}.
The non-smoothness of empirical lower partial moments stated by \citet[Remark~1]{yao2021general} concerns only the second derivatives when the order is two.
For $\SemiVolP$, Lemma~\ref{lem:sv-differentiable} requires only Assumption~\ref{ass:nondeg}, with no continuity assumption on the return distribution, and therefore covers the empirical distributions of Section~\ref{sec:empirical}.

\begin{lemma}[Differentiability of portfolio semivolatility]
\label{lem:sv-differentiable}
Let $\mathcal{W}:=\{x\in\R^n:\SemiVolP(x)>0\}$.
Then $\mathcal{W}$ is open, $\SemiVolP$ is continuously differentiable on $\mathcal{W}$, and
\begin{equation}\label{eq:sv-gradient}
\frac{\partial \SemiVolP(x)}{\partial x_i}
=-\,\frac{\E\!\left[(R_i-\mu_i)\,\bigl((R-\mu)^\top x\bigr)_-\right]}
{\SemiVolP(x)},
\qquad i=1,\dots,n.
\end{equation}
Under Assumption~\ref{ass:nondeg}, $\R^n_+\setminus\{0\}\subseteq\mathcal{W}$, so $\SemiVolP$ is continuously differentiable at every non-trivial long-only portfolio.
\end{lemma}

\begin{proof}
See~\ref{app:proofs}.
\end{proof}

Lemma~\ref{lem:sv-differentiable} corresponds to the differentiability result of \citet[Prop.~4.8 and Cor.~4.9]{fischer2003risk} for the one-sided-moment risk measures $\rho_{p,a}$, $1<p<\infty$.
Under Assumption~\ref{ass:nondeg}, Problem~\eqref{prob:rp-semivol} has a unique optimal solution, by strict concavity of the logarithmic term (see~\ref{app:proofs}), as in \citet{spinu2013,bellini2021risk,ararat2021mad} and, for general risk measures, \citet[Theorems~1 and~2]{cetingoz2024risk}.

The downside risk contributions are
\[
\RC_i(x):=x_i\,\frac{\partial \SemiVolP(x)}{\partial x_i}, \qquad i=1,\dots,n,
\]
well defined on $\R^n_+\setminus\{0\}$ by Lemma~\ref{lem:sv-differentiable}.
Semivariance would give contributions adding up to $2\,\SemiVolP^2$ and the same portfolio, since replacing $\SemiVolP$ by $\SemiVolP^2$ in~\eqref{prob:rp-semivol} leaves the minimizer unchanged.

\begin{proposition}[Risk-budgeting characterization]\label{prop:rp-budget}
Let Assumption~\ref{ass:nondeg} hold and let $y^\star$ be an optimal solution of Problem~\eqref{prob:rp-semivol}.
The logarithmic constraint is active at $y^\star$, and $x^\star := y^\star/\one^\top y^\star$, which does not depend on $\gamma$, is the only $x\in\Delta$ such that
\[
\RC_i(x^\star) = b_i\,\SemiVolP(x^\star), \qquad i=1,\dots,n.
\]
\end{proposition}

\begin{proof}
See \ref{app:proofs}.
\end{proof}

We call $x^{\mathrm{RP}}:=x^\star$ the SV-RP portfolio (strategy RP-SVol$_{\mu}$ of Section~\ref{sec:empirical}); the volatility-based benchmark is recovered by replacing $\SemiVolP$ with $\sigma_P$.\footnote{The implementation of Section~\ref{sec:empirical} solves the penalized form $\min_{y>0}\ \SemiVolP(y)-\sum_i b_i\log(y_i)$.
Under Assumption~\ref{ass:nondeg} its objective is strictly convex and coercive on $\R^n_{++}$, so it has a unique minimizer $\hat y$, which satisfies the KKT conditions of Problem~\eqref{prob:rp-semivol} with multiplier $1$ and $\gamma=\sum_i b_i\log(\hat y_i)$, so that normalization yields the same $x^{\mathrm{RP}}$.
All the risk-budgeting strategies of Section~\ref{sec:empirical} use equal budgets, $b_i=1/n$.}

Under the additivity condition of Remark~\ref{rem:additivity}, $x^{\mathrm{RP}}$ is explicit, with weights proportional to $b_i\,\SemiVol_\mu(R_i)^{-1}$ (Proposition~\ref{prop:svrp-closed-form} of~\ref{app:additivity}).
For $b_i=1/n$ it is the \emph{naive} semivolatility risk-parity portfolio, the downside analogue of the inverse-volatility portfolio \citep{maillard2010properties}.

The results up to Proposition~\ref{prop:rp-budget} and the footnote hold for the threshold-based strategies RP-SVol$_0$ and RP-SVol$_{R^I}$, with $\SemiVolP$ replaced by $\rho_\tau$ in~\eqref{eq:rho-tau}, $R-\mu$ by $R-\tau\one$ in~\eqref{eq:sv-gradient}, and Assumption~\ref{ass:nondeg} by Assumption~\ref{ass:nondeg-tau}, since the proofs use only convexity, degree-one homogeneity and positivity on $\R^n_+\setminus\{0\}$ of the risk functional, and, for the gradient, square integrability, not centering.

\subsection{A reduction lemma for degree-one ratio models}\label{subsec:schaible}

The remaining models maximize the ratio between a linear reward and a convex downside risk functional, a concave--convex fractional program that can be reformulated through the transformation of \citet{schaible1974parameter}.
The reduction is known for reward--risk ratios with a positively homogeneous convex risk functional \citep[Props.~4 and~5]{stoyanov2007optimal} and for diversification ratios \citep{cesarone2025return}.
We restate it in the form used below, with positivity required only on a cone.

\begin{lemma}[Schaible reduction]\label{lem:schaible}
Let \(C\subseteq\R^n_+\) be a closed convex cone, let \(\rho:\R^n_+\to\R_+\) be continuous, convex, positively homogeneous of degree one, and such that \(\rho(x)>0\) for every \(x\in C\setminus\{0\}\), and let \(c\in\R^n\) be such that
\[
\Delta_C^+:=\{x\in C\cap\Delta:\ c^\top x>0\}\neq\emptyset .
\]
Then the fractional problem
\begin{equation}\label{prob:ratio-generic}
\max\left\{\frac{c^\top x}{\rho(x)}\ :\ x\in\Delta_C^+\right\}
\end{equation}
is equivalent to the convex program
\begin{equation}\label{prob:ratio-transf}
\min\left\{\rho(y)\ :\ y\in C,\ c^\top y=1\right\},
\end{equation}
whose minimum is attained.
Moreover, if \(y^\star\) solves~\eqref{prob:ratio-transf}, then \(x^\star=y^\star/(\one^\top y^\star)\) solves~\eqref{prob:ratio-generic}.
\end{lemma}

\begin{proof}
See~\ref{app:proofs}.
\end{proof}

With \(\rho=\sigma_P\), positive away from the origin by Proposition~\ref{prop:nondeg-equiv}(iv), the lemma gives the classical reformulations used for the volatility benchmarks of Section~\ref{sec:empirical}.

\subsection{Most-diversified portfolios under semivolatility}\label{subsec:mdp}

The diversification ratio and the most-diversified portfolio were introduced by \citet{choueifaty2008toward} for volatility and extended to general risk measures by \citet{cesarone2025return}.

Define the asset-level semivolatility coefficients
\[
s_i:=\SemiVol_\mu(R_i) =\left(\E[(R_i-\E[R_i])_-^2]\right)^{1/2}, \qquad i=1,\dots,n,
\]
and the vector \(s:=(s_1,\dots,s_n)^\top\in\R^n_+\).
For \(x \in \Delta\) with \(\SemiVolP(x)>0\), the semivolatility diversification ratio is
\begin{equation}\label{eq:dr}
\DR(x):=
\frac{s^\top x}{\SemiVolP(x)}.
\end{equation}
By subadditivity and degree-one homogeneity of $\SemiVol_\mu$ (Proposition~\ref{prop:semivolatility-properties}), $\SemiVolP(x)\le s^\top x$, hence $\DR(x)\ge1$, as for the volatility-based ratio of \citet{choueifaty2008toward}.

Equality \(\DR(x)=1\) \emph{for every} $x\in\Delta$ is equivalent to additivity of $\SemiVol_\mu$ on the asset returns (Theorem~\ref{thm:sv-additivity-multi}), i.e.\ to a common downside factor for all asset semivolatilities.

Given a minimum expected return \(\eta\), a most-diversified portfolio under semivolatility solves
\begin{equation}\label{prob:mdp-direct}
\begin{aligned}
\max_{x \in \R^n}\quad & \DR(x) \\
\text{s.t.}\quad
& \mu^\top x \ge \eta,\\
& \one^\top x = 1,\\
& x \ge 0.
\end{aligned}
\end{equation}
Lemma~\ref{lem:schaible} applies with \(\rho=\SemiVolP\), \(c=s\) and
\[
C:=\{y\in\R^n_+ : \mu^\top y \ge \eta\,\one^\top y\},
\]
a closed convex cone on which \(\rho>0\) away from the origin under Assumption~\ref{ass:nondeg}, and, since \(s_i>0\) under the same assumption, \(\Delta_C^+\neq\emptyset\) amounts to \(\eta\le\max_i\mu_i\).
Problem~\eqref{prob:mdp-direct} is therefore equivalent to the convex program
\begin{equation}\label{prob:mdp-schaible}
\begin{aligned}
\min_{y \in \R^n}\quad & \SemiVolP(y) \\
\text{s.t.}\quad
& s^\top y = 1,\\
& \mu^\top y \ge \eta\,\one^\top y,\\
& y \ge 0,
\end{aligned}
\end{equation}
with a most-diversified portfolio $x^{\mathrm{MD}}=y^\star/\one^\top y^\star$ for every solution $y^\star$.
Replacing \(s_i\) and \(\SemiVolP\) with \(\sd(R_i)\) and \(\sigma_P\) recovers the volatility-based version.
Replacing \(s_i\) and \(\SemiVolP\) with \(\rho_\tau(e_i)=\SemiVol_\tau(R_i)\) and \(\rho_\tau\) gives MDP-SVol$_0$ and MDP-SVol$_{R^I}$, for which \(\DR\ge1\) by subadditivity of \(\rho_\tau\) and Lemma~\ref{lem:schaible} applies under Assumption~\ref{ass:nondeg-tau}.

\citet{barkhagen2023optimising} consider ratios of convex functions homogeneous of equal degree, such as kurtosis, which are in general not quasi-concave and need global algorithms, whereas for convex degree-one functionals Lemma~\ref{lem:schaible} avoids that machinery.

\subsection{Sharpe versus Sortino ratio optimization}\label{subsec:sortino}

The classical Sharpe-ratio problem is
\begin{equation}\label{prob:sharpe}
\begin{aligned}
\max_{x \in \R^n}\quad &
\Sharpe(x):=
\frac{\mu^\top x-r_f}{\sigma_P(x)}\\
\text{s.t.}\quad
& \one^\top x = 1,\\
& x \ge 0,
\end{aligned}
\end{equation}
with \(r_f\) the risk-free rate over the horizon.
The downside analogue is the Sortino ratio \citep{sortino1991downside,sortino1994performance}, with a target $\tau\in L^2$, deterministic or random, as in the outperformance ratio of \citet{meade2011detection}.
On the simplex $R_P(x)-\tau=(R-\tau\one)^\top x$, and the Sortino ratio reads
\begin{equation}\label{eq:sortino-target}
\Sortino_\tau(x):=
\frac{\E[R_P(x)-\tau]}{\SemiVol_\tau(R_P(x))}
=\frac{\bigl(\mu-\E[\tau]\one\bigr)^\top x}{\rho_\tau(x)} .
\end{equation}
For $\tau=R^I$ the numerator is the expected active return $(\mu-\E[R^I]\one)^\top x$, so that reward and risk are both benchmark-relative.

We also consider the below-mean counterpart of~\eqref{eq:sortino-target}, in which the portfolio mean replaces the target in the semideviation while the numerator keeps the excess return over $r_f$,
\begin{equation}\label{eq:sortino-mean}
\Sortino_\mu(x):=
\frac{\mu^\top x-r_f}{\SemiVolP(x)}
=
\frac{\mu^\top x-r_f}{\SemiVol_\mu(R_P(x))},
\end{equation}
whose denominator is the below-mean semivolatility of the risk-parity and most-diversified models of Sections~\ref{subsec:rp} and~\ref{subsec:mdp}.
The denominator $\SemiVol_\mu$ is a deviation measure \citep{rockafellar2006generalized}, whereas $\SemiVol_\tau$ measures shortfall against an exogenous target.
For deterministic $\tau$ and non-constant $R_P(x)$ the two coincide only when $\mu^\top x=\tau$.

Setting \(\widetilde{\mu}:=\mu-r_f\one\) for~\eqref{eq:sortino-mean} and \(\widetilde{\mu}:=\mu-\E[\tau]\one\) for~\eqref{eq:sortino-target}, both ratios are maximized by solving
\begin{equation}\label{prob:sortino-mean}
\begin{aligned}
\max_{x \in \R^n}\quad &
\frac{\widetilde{\mu}^\top x}{\rho(x)}\\
\text{s.t.}\quad
& \one^\top x = 1,\\
& x \ge 0,
\end{aligned}
\end{equation}
with $\rho=\SemiVolP$ and $\rho=\rho_\tau$, respectively.
Lemma~\ref{lem:schaible} applies to both with $C=\R^n_+$ and $c=\widetilde\mu$, positivity holding under Assumption~\ref{ass:nondeg} by Proposition~\ref{prop:nondeg-equiv}(iii) and under Assumption~\ref{ass:nondeg-tau}, respectively.
Hence, whenever $\widetilde\mu_i>0$ for some $i$, Problem~\eqref{prob:sortino-mean} restricted to $\Delta^+:=\{x\in\Delta:\widetilde\mu^\top x>0\}$ is equivalent to the convex program
\begin{equation}\label{prob:sortino-mean-transf}
\begin{aligned}
\min_{y \in \R^n}\quad & \rho(y) \\
\text{s.t.}\quad
& \widetilde{\mu}^\top y = 1,\\
& y \ge 0,
\end{aligned}
\end{equation}
with optimal portfolio $x^\star=y^\star/(\one^\top y^\star)$.
The restriction loses nothing, since the optimal ratio over $\Delta$ is then positive and attained only in $\Delta^+$.
Under Assumptions~\ref{ass:nondeg-tau} and~\ref{ass:nondeg}, respectively, the target-based ratios~\eqref{eq:sortino-target} and the below-mean ratio~\eqref{eq:sortino-mean} thus lead to convex programs of the same form~\eqref{prob:sortino-mean-transf}, with $\rho=\rho_\tau$ and $\rho=\SemiVolP$ and the corresponding $\widetilde\mu$, so that the threshold is a modeling, not a computational, choice.

The exact solution of Omega-ratio maximization through a fractional linear program \citep{kapsos2014optimizing} is also an instance of Lemma~\ref{lem:schaible}.
Indeed, for deterministic $\tau$, on $\Delta$ the Omega ratio satisfies $\E[(R_P(x)-\tau)_+]/\E[(R_P(x)-\tau)_-]-1 =(\mu-\tau\one)^\top x/\E\bigl[\bigl((R-\tau\one)^\top x\bigr)_-\bigr]$ \citep{kaplan2004kappa}, and the lemma applies under Assumption~\ref{ass:nondeg-tau} whenever $\mu_i>\tau$ for some $i$.

\section{Empirical analysis}\label{sec:empirical}

\subsection{Datasets and experimental setup}\label{subsec:data}

The empirical analysis is based on two datasets obtained from FactSet \citep{FactSet} and listed in Table~\ref{tab:datasets}.
They consist of the daily total-return prices of the constituents of two US equity indexes, adjusted for stock splits, dividends and spin-offs, together with the levels of the total-return versions of the indexes.
The data were downloaded on November 6, 2026, and are available with the results at \href{https://www.francescocesarone.com/data-sets}{francescocesarone.com} \citep{Cesarone2026data}, subject to the license terms of the data provider.

\begin{table}[htbp]
\centering
\caption{Datasets analyzed; $n$: assets with a complete price series.}
\label{tab:datasets}
\small
\renewcommand{\arraystretch}{1.15}
\makebox[\textwidth][c]{%
\begin{adjustbox}{width=1.45\textwidth}
\begin{tabular}{>{\raggedright\arraybackslash}p{5cm}lccl}
\toprule
\textbf{Index}
& \textbf{Full name}
& $\bm{n}$
& \textbf{Country}
& \textbf{Time interval} \\
\midrule
DowJones   & Dow Jones Industrial Average            & 30  & USA & Mar 2008\,--\,Nov 2026 \\
Nasdaq100  & Nasdaq-100                              & 84  & USA & Oct 2012\,--\,Nov 2026 \\
\bottomrule
\end{tabular}
\end{adjustbox}%
}
\end{table}

Asset returns are computed as simple returns, namely $r_{it} = (P_{it} - P_{i,t-1})/P_{i,t-1}$, where $P_{it}$ denotes the adjusted closing price of asset~$i$ at time~$t$, and we write $r_t=(r_{1t},\dots,r_{nt})^\top$.
The index return $R^I_t$ is computed in the same way from the level of the total-return index.

We evaluate the selected portfolios out of sample with a rolling time-window approach \citep{demiguel2009optimal,kritzman2010defense,cesarone2017minimum,cesarone2025return,cesarone2025semivariance}.
Let $T$ denote the number of daily observations, and let $D_L = 504$ trading days (about two years) and $D_H = 21$ trading days (one month) be the lengths of the \emph{learning} (in-sample) and \emph{holding} (out-of-sample) periods.
The rolling procedure generates $Q = \lfloor (T - D_L) / D_H \rfloor$ windows, indexed by $q = 1, \ldots, Q$, where the in-sample period of window~$q$ ends at day $t^{\mathrm{end}}_q = D_H(q-1) + D_L$.
In each window every model is estimated and solved on the in-sample data, giving the optimal allocation $x^{\star}_q$, which is held without trading over the subsequent holding period, so that the weights drift with prices.
The out-of-sample portfolio return of day $t$ is then
\[
R^{\mathrm{out}}_t = \sum_{i=1}^{n} x_{t-1,i}\, r_{i,t},
\qquad
x_{t-1,i} = \frac{x^{\star}_{q,i} \prod_{s=t^{\mathrm{end}}_q+1}^{t-1} (1 + r_{i,s})}{\sum_{j=1}^{n} x^{\star}_{q,j} \prod_{s=t^{\mathrm{end}}_q+1}^{t-1} (1 + r_{j,s})},
\]
for $t = t^{\mathrm{end}}_q + 1, \ldots, t^{\mathrm{end}}_q + D_H$, where $x_{t-1,i}$ is the weight of asset~$i$ at the close of day $t-1$.
Concatenating the holding-period returns gives $N^{\mathrm{out}} = Q\,D_H$ out-of-sample observations, with $Q=199$ windows from about March 2010 on the DowJones and $Q=144$ windows from about October 2014 on the Nasdaq100, both ending in October 2026.

The investment universe is static and consists, for each dataset, of the constituents of the index at the download date whose price series is complete over the whole time interval considered.
This rule retains the $n$ assets of Table~\ref{tab:datasets} out of $30$ and $101$ constituents respectively.
For the Nasdaq100, FER-US, whose price is unchanged on most days until 2020, is excluded by hand.
Such a universe is subject to survivorship and look-ahead bias \citep{brown1992survivorship,elton1996survivor}, which favors every strategy relative to the market index.
We therefore use the index only as a reference and draw our conclusions from comparisons among strategies.
The results of the four strategies whose threshold is the index return should be read with particular caution, since the index enters their optimization itself.

Following \citet{demiguel2009optimal}, at each rebalancing date we charge proportional transaction costs of 50 basis points on the portfolio turnover.
With $x^{+}_{q-1}$ the weights of window $q-1$ drifted to the close of day $t^{\mathrm{end}}_q$, the return $R^{\mathrm{out}}_t$ of day $t=t^{\mathrm{end}}_q+1$ is reduced by $c_{\mathrm{tc}} \sum_{i=1}^{n} |x^{\star}_{q,i} - x^{+}_{q-1,i}|$, with $c_{\mathrm{tc}} = 0.0050$.
No cost is charged on the initial allocation of the first window.
The Supplementary Material also reports the Sharpe ratios at $0$, $25$ and $100$ basis points.

All the experiments have been implemented in MATLAB R2025b.
The quadratic programs of MV, MaxSharpe and MDP-Vol are solved with \texttt{quadprog} (interior-point-convex) and the other models with \texttt{fmincon} (sequential quadratic programming with analytic gradients, started from the equally weighted portfolio, or the inverse-semivolatility one for MDP-SVol, scaled to feasibility).
The tolerances are $10^{-12}$ for the quadratic and ratio programs and $10^{-10}$ for the others.
In the homogeneous reformulations of the ratio models, a cap $x_i\le\mathrm{ub}$ on the normalized weights is imposed as the linear constraint $y_i\le\mathrm{ub}\,\one^\top y$, which keeps the feasible set a closed convex cone and is exact after normalization; it is redundant at the baseline $\mathrm{ub}=1$, and the risk-budgeting programs carry only the bound $y_i\ge10^{-10}$.

\subsection{Portfolio models under comparison}\label{subsec:strategies}

We compare the out-of-sample performance of seventeen long-only portfolio strategies ($x_i \ge 0$, $\one^\top x = 1$, with no upper bound on the individual weights).
Apart from the Equally Weighted (EW) portfolio of \citet{demiguel2009optimal}, they form a grid along the three axes of Table~\ref{tab:strategies}, namely the \emph{model class}, the \emph{risk measure} and the \emph{threshold} below which deviations are measured.
The threshold is the portfolio mean, the fixed level $\tau=0$ or the market-index return $\tau_t=R^{I}_t$.
The market index is reported as an additional row in the tables of Section~\ref{sec:results}, as an external reference rather than a strategy, so the entries that require portfolio weights, the turnover and the number of assets, are marked as not applicable (---) for it.

\begin{table}[htbp]
\centering
\caption{The seventeen portfolio models under comparison.}
\label{tab:strategies}
\small
\renewcommand{\arraystretch}{1.15}
\begin{adjustbox}{max width=\textwidth}
\begin{tabular}{lllc}
\toprule
\textbf{Strategy} & \textbf{Model class} & \textbf{Risk measure} &
$\bm{\tau}$ \\
\midrule
EW & naive & --- & --- \\
\hdashline
MV            & \multirow{4}{*}{minimum risk} & variance     & --- \\
MSV$_{0}$     &                               & semivariance & $0$ \\
MSV$_{\mu}$   &                               & semivariance & $\mu^\top x$ \\
MSV$_{R^{I}}$ &                               & semivariance & $R^{I}_t$ \\
\hdashline
MaxSharpe            & \multirow{4}{*}{reward-to-risk ratio} & volatility     & --- \\
MaxSortino$_{\mu}$   &                                       & semivolatility & $\mu^\top x$ \\
MaxSortino$_{0}$     &                                       & semivolatility & $0$ \\
MaxSortino$_{R^{I}}$ &                                       & semivolatility & $R^{I}_t$ \\
\hdashline
RP-Vol            & \multirow{4}{*}{risk budgeting} & volatility     & --- \\
RP-SVol$_{\mu}$   &                                 & semivolatility & $\mu^\top x$ \\
RP-SVol$_{0}$     &                                 & semivolatility & $0$ \\
RP-SVol$_{R^{I}}$ &                                 & semivolatility & $R^{I}_t$ \\
\hdashline
MDP-Vol            & \multirow{4}{*}{diversification ratio} & volatility     & --- \\
MDP-SVol$_{\mu}$   &                                        & semivolatility & $\mu^\top x$ \\
MDP-SVol$_{0}$     &                                        & semivolatility & $0$ \\
MDP-SVol$_{R^{I}}$ &                                        & semivolatility & $R^{I}_t$ \\
\bottomrule
\end{tabular}
\end{adjustbox}
\end{table}

In any window in which the solver returns a non-positive exit flag or stops with an error, the model falls back to the equally weighted allocation.
In the baseline configuration this fallback is never needed, since every model returns a positive exit flag in every rolling window of both universes.
In the accepted solutions, negative entries at the tolerance level are set to zero and the weights are renormalized.
In every window the reward-to-risk ratio models satisfy $\Delta_C^+\neq\emptyset$ (some asset has a positive in-sample mean return and some asset an in-sample mean return above that of the index), and Assumption~\ref{ass:nondeg-tau} holds for $\tau=0$ and $\tau_t=R^I_t$.
The assumption is verified through the linear program $m^\star=\min_{x\in\Delta}\max_t(\tau_t-r_t^\top x)$ on the sample returns, whose value is positive if and only if Assumption~\ref{ass:nondeg-tau} holds in that window.
The accuracy of the risk-budgeting solutions is documented in Section~S.7.2 of the Supplementary Material.

The risk-free rate is set to $r_f = 0$, both in the objectives of MaxSharpe and MaxSortino$_{\mu}$ and in the performance measures of Section~\ref{subsec:measures}, as in \citet{cesarone2017minimum}.

\subsection{Performance measures}\label{subsec:measures}

The portfolio performance is assessed through the following measures, typically adopted in portfolio selection (see, e.g., \citealt{demiguel2009optimal,cesarone2017minimum,cesarone2025return}, and references therein).
All the measures are computed on the concatenated out-of-sample portfolio return series, re-indexed as $R^{\mathrm{out}}_t$, $t=1,\dots,N^{\mathrm{out}}$, net of transaction costs.
We denote by $W_t = W_{t-1}\bigl(1+R^{\mathrm{out}}_t\bigr)$ the portfolio wealth at time $t$, with $W_0 = 1$, and by
\begin{equation}\label{eq:drawdown}
\dd_t := \frac{W_t-\max_{0\le \nu\le t}W_\nu}{\max_{0\le \nu\le t}W_\nu},
\qquad t=1,\dots,N^{\mathrm{out}},
\end{equation}
the corresponding drawdown path.
Higher values are preferable for $\mu^{\mathrm{ann}}$, Sharpe, Sortino, Omega, Rachev and Calmar, lower values for $\sigma^{\mathrm{ann}}$, CVaR$_{5\%}$, Maximum Drawdown, Ulcer index and Turnover.
\begin{itemize}[leftmargin=2em]
  \item The annualized average of the out-of-sample portfolio returns,
  \begin{equation}\label{eq:muann}
  \mu^{\mathrm{ann}} = \frac{252}{N^{\mathrm{out}}}
  \sum_{t=1}^{N^{\mathrm{out}}} R^{\mathrm{out}}_t .
  \end{equation}
  \item The annualized standard deviation of the out-of-sample portfolio
  returns,
  \begin{equation}\label{eq:sigmaann}
  \sigma^{\mathrm{ann}} = \sqrt{252}\,
  \left(\frac{1}{N^{\mathrm{out}}-1}\sum_{t=1}^{N^{\mathrm{out}}}
  \bigl(R^{\mathrm{out}}_t-\bar R^{\mathrm{out}}\bigr)^2\right)^{1/2},
  \end{equation}
  where $\bar R^{\mathrm{out}} := \frac{1}{N^{\mathrm{out}}}\sum_{t=1}^{N^{\mathrm{out}}} R^{\mathrm{out}}_t$.
  \item The Sharpe ratio, the gain per unit of total risk assumed,
  \begin{equation}\label{eq:sharpe-out}
  \Sharpe^{\mathrm{out}} =
  \frac{\mu^{\mathrm{ann}}-r_f}{\sigma^{\mathrm{ann}}} ,
  \end{equation}
  where $r_f$ is here the annual counterpart of the rate of Section~\ref{subsec:sortino}; in~\eqref{eq:sortino-out} and~\eqref{eq:rachev}, which are computed on daily returns, it is replaced by the daily rate $r_f^{\mathrm{d}}:=r_f/252$.
  \item The Sortino ratio \citep{sortino1991downside,sortino1994performance}
  represents a variation of the Sharpe ratio, where, instead of the standard
  deviation, the downside deviation of the out-of-sample returns below the
  target $r_f^{\mathrm{d}}$ is considered:
  \begin{equation}\label{eq:sortino-out}
  \Sortino^{\mathrm{out}} = \sqrt{252}\,
  \frac{\bar R^{\mathrm{out}}-r_f^{\mathrm{d}}}
       {\left(\frac{1}{N^{\mathrm{out}}}\sum_{t=1}^{N^{\mathrm{out}}}
       \bigl(R^{\mathrm{out}}_t-r_f^{\mathrm{d}}\bigr)_-^2\right)^{1/2}} .
  \end{equation}
  The same target is used for every strategy, whatever the threshold of its objective.
  \item The Omega ratio \citep{keating2002universal} is the ratio of the
  expected gains above a threshold $\tau$ to the expected losses below it,
  \begin{equation}\label{eq:omega}
  \Omega_\tau = \frac{\sum_{t=1}^{N^{\mathrm{out}}}
  \bigl(R^{\mathrm{out}}_t-\tau\bigr)_+}
  {\sum_{t=1}^{N^{\mathrm{out}}}\bigl(R^{\mathrm{out}}_t-\tau\bigr)_-},
  \qquad \tau = 0 .
  \end{equation}
  \item The Rachev ratio \citep{biglova2004different} measures the upside
  potential by comparing the right and the left tails of the out-of-sample
  returns,
  \begin{equation}\label{eq:rachev}
  \Rachev_\alpha =
  \frac{\CVaR_\alpha\bigl(r_f^{\mathrm{d}}-R^{\mathrm{out}}\bigr)}
       {\CVaR_\alpha\bigl(R^{\mathrm{out}}-r_f^{\mathrm{d}}\bigr)},
  \qquad \alpha = 10\% ,
  \end{equation}
  where $\CVaR_\alpha(\cdot)$ is the Conditional Value-at-Risk of~\eqref{eq:cvar} applied to the variable in parentheses, computed on the $\lceil\alpha N^{\mathrm{out}}\rceil$ largest, respectively smallest, returns.
  \item The Conditional Value-at-Risk \citep{rockafellar2000optimization} at the
  $\varepsilon$ level,
  \begin{equation}\label{eq:cvar}
  \CVaR_\varepsilon\bigl(R^{\mathrm{out}}\bigr) = -\,\E\!\left[R^{\mathrm{out}}\mid R^{\mathrm{out}}\le
  q_\varepsilon\bigl(R^{\mathrm{out}}\bigr)\right],
  \qquad \varepsilon = 5\% ,
  \end{equation}
  where $q_\varepsilon$ is the empirical $\varepsilon$-quantile, is the average
  of the $5\%$ worst out-of-sample returns, with sign changed.
  \item The Maximum Drawdown measures the maximal
  out-of-sample loss from the observed peak,
  \begin{equation}\label{eq:mdd}
  \MaxDD = \Bigl|\min_{1\le t\le N^{\mathrm{out}}} \dd_t\Bigr| .
  \end{equation}
  \item The Calmar ratio \citep{young1991calmar} measures the gain per unit of
  maximal loss suffered, and is computed as
  \begin{equation}\label{eq:calmar}
  \Calmar = \frac{\mu^{\mathrm{ann}}-r_f}{\MaxDD} .
  \end{equation}
  \item The Ulcer index \citep{martin1989investor} evaluates the depth and the
  duration of the drawdowns,
  \begin{equation}\label{eq:ulcer}
  \UI = \left(\frac{1}{N^{\mathrm{out}}+1}
  \sum_{t=0}^{N^{\mathrm{out}}} \dd_t^2\right)^{1/2} ,
  \end{equation}
  the average running over the whole wealth path, its initial point included ($\dd_0=0$).
  \item The Turnover \citep{demiguel2009optimal} evaluates the amount of trading
  required to implement the portfolio strategy in practice,
  \begin{equation}\label{eq:turnover}
  \Turn = \frac{1}{Q-1}\sum_{q=2}^{Q}\sum_{i=1}^{n}
  \bigl|x^{\star}_{q,i}-x^{+}_{q-1,i}\bigr| ,
  \end{equation}
  with $x^{+}_{q-1}$ as above, and is a proxy for transaction costs.
  \item The average number of assets held in the portfolio,
  \begin{equation}\label{eq:nassets}
  \#\mathrm{Assets} = \frac{1}{Q}\sum_{q=1}^{Q}
  \Bigl|\bigl\{i : x^{\star}_{q,i} > 10^{-6}\bigr\}\Bigr| .
  \end{equation}
  Being a descriptive indicator rather than a performance criterion, it is
  neither ranked nor highlighted in the tables.
\end{itemize}

\subsection{Computational results}\label{sec:results}

Tables~\ref{tab:oos_dowjones} and~\ref{tab:oos_nasdaq} report the out-of-sample performance results for the seventeen portfolio strategies on the DowJones and Nasdaq100 datasets, respectively.
For each measure the best value is in bold and the cells are colored by rank, from deep green for the best to deep red for the worst, the market-index row in last position being excluded from both.
We first describe each universe in turn; Section~\ref{subsubsec:all} tests the within-family differences along the risk-measure and threshold axes of Table~\ref{tab:strategies} and gathers what survives across the two datasets.

\subsubsection{DowJones}\label{subsubsec:dowjones}

\begin{table}[htbp]
\centering
\caption{Out-of-sample performance results for the DowJones dataset. Ratios (Sharpe, Sortino, Calmar, Omega, Rachev$_{10\%}$) in natural units, three decimals; all other values except \#Assets in percent, two decimals. CVaR$_{5\%}$ is computed on daily returns and Turnover per rebalancing.}
\label{tab:oos_dowjones}
\small
\renewcommand{\arraystretch}{1.15}
\begin{adjustbox}{width=\textwidth}
\begin{tabular}{lcccccccccccc}
\toprule
\textbf{Model} & $\mu^{\mathrm{ann}}$ & $\sigma^{\mathrm{ann}}$ & Sharpe & Sortino & Calmar & Omega & Rachev$_{10\%}$ & CVaR$_{5\%}$ & MaxDD & Ulcer & Turn & \#Assets \\
\midrule
EW & \cellcolor[RGB]{158,207,126}17.15 & \cellcolor[RGB]{254,219,129}16.34 & \cellcolor[RGB]{138,201,125}1.049 & \cellcolor[RGB]{138,201,125}1.499 & \cellcolor[RGB]{99,190,123}\textbf{0.529} & \cellcolor[RGB]{118,196,124}1.222 & \cellcolor[RGB]{253,194,124}0.979 & \cellcolor[RGB]{252,170,120}2.45 & \cellcolor[RGB]{254,219,129}32.42 & \cellcolor[RGB]{196,218,129}4.92 & \cellcolor[RGB]{118,196,124}4.26 & 30.00 \\
\midrule
MV & \cellcolor[RGB]{249,121,110}10.07 & \cellcolor[RGB]{138,201,125}13.01 & \cellcolor[RGB]{249,121,110}0.774 & \cellcolor[RGB]{249,121,110}1.101 & \cellcolor[RGB]{248,105,107}0.340 & \cellcolor[RGB]{249,121,110}1.156 & \cellcolor[RGB]{196,218,129}0.984 & \cellcolor[RGB]{118,196,124}\textbf{1.88} & \cellcolor[RGB]{138,201,125}29.64 & \cellcolor[RGB]{254,219,129}5.29 & \cellcolor[RGB]{236,229,131}12.73 & 12.38 \\
MSV$_{0}$ & \cellcolor[RGB]{250,138,113}10.21 & \cellcolor[RGB]{118,196,124}12.96 & \cellcolor[RGB]{250,138,113}0.787 & \cellcolor[RGB]{250,138,113}1.119 & \cellcolor[RGB]{250,138,113}0.378 & \cellcolor[RGB]{250,138,113}1.157 & \cellcolor[RGB]{177,212,128}0.985 & \cellcolor[RGB]{118,196,124}\textbf{1.88} & \cellcolor[RGB]{99,190,123}\textbf{26.99} & \cellcolor[RGB]{252,170,120}5.57 & \cellcolor[RGB]{252,186,123}15.48 & 11.45 \\
MSV$_{\mu}$ & \cellcolor[RGB]{248,105,107}9.93 & \cellcolor[RGB]{99,190,123}\textbf{12.95} & \cellcolor[RGB]{248,105,107}0.767 & \cellcolor[RGB]{248,105,107}1.089 & \cellcolor[RGB]{249,121,110}0.355 & \cellcolor[RGB]{248,105,107}1.153 & \cellcolor[RGB]{158,207,126}0.991 & \cellcolor[RGB]{118,196,124}\textbf{1.88} & \cellcolor[RGB]{118,196,124}27.96 & \cellcolor[RGB]{251,154,116}5.76 & \cellcolor[RGB]{254,219,129}14.04 & 11.51 \\
MSV$_{R^{I}}$ & \cellcolor[RGB]{251,162,118}14.34 & \cellcolor[RGB]{252,186,123}16.44 & \cellcolor[RGB]{253,194,124}0.872 & \cellcolor[RGB]{252,170,120}1.238 & \cellcolor[RGB]{251,162,118}0.385 & \cellcolor[RGB]{254,211,127}1.184 & \cellcolor[RGB]{255,227,130}0.980 & \cellcolor[RGB]{254,219,129}2.43 & \cellcolor[RGB]{251,154,116}37.24 & \cellcolor[RGB]{245,232,131}5.20 & \cellcolor[RGB]{196,218,129}5.75 & 28.25 \\
\midrule
MaxSharpe & \cellcolor[RGB]{138,201,125}18.37 & \cellcolor[RGB]{250,138,113}20.76 & \cellcolor[RGB]{254,219,129}0.885 & \cellcolor[RGB]{236,229,131}1.308 & \cellcolor[RGB]{216,224,130}0.489 & \cellcolor[RGB]{252,186,123}1.178 & \cellcolor[RGB]{138,201,125}1.011 & \cellcolor[RGB]{250,138,113}3.03 & \cellcolor[RGB]{250,138,113}37.58 & \cellcolor[RGB]{250,138,113}11.93 & \cellcolor[RGB]{250,138,113}42.15 & 6.84 \\
MaxSortino$_{\mu}$ & \cellcolor[RGB]{118,196,124}19.20 & \cellcolor[RGB]{249,121,110}22.32 & \cellcolor[RGB]{252,170,120}0.860 & \cellcolor[RGB]{255,235,132}1.289 & \cellcolor[RGB]{254,219,129}0.477 & \cellcolor[RGB]{252,170,120}1.176 & \cellcolor[RGB]{118,196,124}1.030 & \cellcolor[RGB]{249,121,110}3.21 & \cellcolor[RGB]{248,105,107}40.26 & \cellcolor[RGB]{249,121,110}13.19 & \cellcolor[RGB]{249,121,110}44.50 & 6.30 \\
MaxSortino$_{0}$ & \cellcolor[RGB]{99,190,123}\textbf{19.23} & \cellcolor[RGB]{248,105,107}22.48 & \cellcolor[RGB]{251,154,116}0.855 & \cellcolor[RGB]{254,219,129}1.282 & \cellcolor[RGB]{255,235,132}0.478 & \cellcolor[RGB]{251,154,116}1.175 & \cellcolor[RGB]{99,190,123}\textbf{1.031} & \cellcolor[RGB]{248,105,107}3.24 & \cellcolor[RGB]{249,121,110}40.24 & \cellcolor[RGB]{248,105,107}13.21 & \cellcolor[RGB]{248,105,107}44.99 & 6.17 \\
MaxSortino$_{R^{I}}$ & \cellcolor[RGB]{253,202,126}14.86 & \cellcolor[RGB]{251,154,116}16.59 & \cellcolor[RGB]{236,229,131}0.896 & \cellcolor[RGB]{253,202,126}1.271 & \cellcolor[RGB]{253,202,126}0.402 & \cellcolor[RGB]{245,232,131}1.188 & \cellcolor[RGB]{248,105,107}0.975 & \cellcolor[RGB]{251,154,116}2.47 & \cellcolor[RGB]{253,202,126}36.95 & \cellcolor[RGB]{253,202,126}5.38 & \cellcolor[RGB]{255,235,132}13.25 & 26.20 \\
\midrule
RP-Vol & \cellcolor[RGB]{216,224,130}15.78 & \cellcolor[RGB]{245,232,131}15.17 & \cellcolor[RGB]{177,212,128}1.040 & \cellcolor[RGB]{167,210,127}1.484 & \cellcolor[RGB]{187,215,128}0.502 & \cellcolor[RGB]{148,204,126}1.220 & \cellcolor[RGB]{251,162,118}0.978 & \cellcolor[RGB]{236,229,131}2.27 & \cellcolor[RGB]{236,229,131}31.41 & \cellcolor[RGB]{138,201,125}4.48 & \cellcolor[RGB]{138,201,125}4.43 & 30.00 \\
RP-SVol$_{\mu}$ & \cellcolor[RGB]{245,232,131}15.73 & \cellcolor[RGB]{216,224,130}15.13 & \cellcolor[RGB]{196,218,129}1.039 & \cellcolor[RGB]{196,218,129}1.482 & \cellcolor[RGB]{187,215,128}0.502 & \cellcolor[RGB]{187,215,128}1.219 & \cellcolor[RGB]{250,138,113}0.977 & \cellcolor[RGB]{236,229,131}2.27 & \cellcolor[RGB]{216,224,130}31.33 & \cellcolor[RGB]{99,190,123}\textbf{4.44} & \cellcolor[RGB]{158,207,126}4.52 & 30.00 \\
RP-SVol$_{0}$ & \cellcolor[RGB]{245,232,131}15.73 & \cellcolor[RGB]{196,218,129}15.12 & \cellcolor[RGB]{158,207,126}1.041 & \cellcolor[RGB]{167,210,127}1.484 & \cellcolor[RGB]{158,207,126}0.504 & \cellcolor[RGB]{148,204,126}1.220 & \cellcolor[RGB]{249,121,110}0.976 & \cellcolor[RGB]{236,229,131}2.27 & \cellcolor[RGB]{196,218,129}31.20 & \cellcolor[RGB]{118,196,124}4.45 & \cellcolor[RGB]{99,190,123}\textbf{4.21} & 30.00 \\
RP-SVol$_{R^{I}}$ & \cellcolor[RGB]{252,186,123}14.63 & \cellcolor[RGB]{253,202,126}16.41 & \cellcolor[RGB]{255,235,132}0.891 & \cellcolor[RGB]{252,186,123}1.265 & \cellcolor[RGB]{252,186,123}0.396 & \cellcolor[RGB]{245,232,131}1.188 & \cellcolor[RGB]{251,162,118}0.978 & \cellcolor[RGB]{253,194,124}2.44 & \cellcolor[RGB]{252,186,123}36.98 & \cellcolor[RGB]{216,224,130}5.18 & \cellcolor[RGB]{177,212,128}4.93 & 30.00 \\
\midrule
MDP-Vol & \cellcolor[RGB]{254,219,129}15.59 & \cellcolor[RGB]{245,232,131}15.17 & \cellcolor[RGB]{216,224,130}1.028 & \cellcolor[RGB]{216,224,130}1.464 & \cellcolor[RGB]{236,229,131}0.483 & \cellcolor[RGB]{216,224,130}1.211 & \cellcolor[RGB]{216,224,130}0.983 & \cellcolor[RGB]{196,218,129}2.26 & \cellcolor[RGB]{255,235,132}32.25 & \cellcolor[RGB]{252,186,123}5.39 & \cellcolor[RGB]{253,202,126}14.36 & 15.98 \\
MDP-SVol$_{\mu}$ & \cellcolor[RGB]{196,218,129}15.82 & \cellcolor[RGB]{177,212,128}14.77 & \cellcolor[RGB]{118,196,124}1.071 & \cellcolor[RGB]{118,196,124}1.528 & \cellcolor[RGB]{138,201,125}0.509 & \cellcolor[RGB]{187,215,128}1.219 & \cellcolor[RGB]{236,229,131}0.982 & \cellcolor[RGB]{167,210,127}2.19 & \cellcolor[RGB]{177,212,128}31.07 & \cellcolor[RGB]{177,212,128}4.52 & \cellcolor[RGB]{251,154,116}17.41 & 15.23 \\
MDP-SVol$_{0}$ & \cellcolor[RGB]{177,212,128}16.06 & \cellcolor[RGB]{158,207,126}14.74 & \cellcolor[RGB]{99,190,123}\textbf{1.089} & \cellcolor[RGB]{99,190,123}\textbf{1.555} & \cellcolor[RGB]{118,196,124}0.525 & \cellcolor[RGB]{99,190,123}\textbf{1.223} & \cellcolor[RGB]{255,227,130}0.980 & \cellcolor[RGB]{167,210,127}2.19 & \cellcolor[RGB]{158,207,126}30.61 & \cellcolor[RGB]{158,207,126}4.49 & \cellcolor[RGB]{252,170,120}17.28 & 15.05 \\
MDP-SVol$_{R^{I}}$ & \cellcolor[RGB]{251,162,118}14.34 & \cellcolor[RGB]{252,170,120}16.45 & \cellcolor[RGB]{253,194,124}0.872 & \cellcolor[RGB]{251,154,116}1.237 & \cellcolor[RGB]{251,162,118}0.385 & \cellcolor[RGB]{254,211,127}1.184 & \cellcolor[RGB]{253,194,124}0.979 & \cellcolor[RGB]{253,194,124}2.44 & \cellcolor[RGB]{252,170,120}37.22 & \cellcolor[RGB]{245,232,131}5.20 & \cellcolor[RGB]{216,224,130}5.77 & 28.62 \\
\midrule
Market index & 13.18 & 16.47 & 0.801 & 1.131 & 0.358 & 1.167 & 0.970 & 2.45 & 36.82 & 5.45 & --- & --- \\
\bottomrule
\end{tabular}
\end{adjustbox}
\end{table}

As shown in Table~\ref{tab:oos_dowjones}, MDP-SVol$_{0}$ and MDP-SVol$_{\mu}$ achieve the highest Sharpe ratios ($1.089$ and $1.071$), ahead of the equally weighted portfolio ($1.049$) and of MDP-Vol ($1.028$).
As expected, the minimum-risk portfolios MV, MSV$_{0}$ and MSV$_{\mu}$ show the lowest volatility, CVaR$_{5\%}$ and Maximum Drawdown, but also the worst annualized returns.
They are the only strategies with a Sharpe ratio below the market index ($0.801$), a comparison that the construction of the universe tilts toward the strategies (Section~\ref{subsec:data}).
Conversely, MaxSharpe, MaxSortino$_{\mu}$ and MaxSortino$_{0}$ provide the highest returns but the worst values of all risk measures and turnover, with the two MaxSortino portfolios slightly below MaxSharpe in Sharpe ratio.
RP-Vol, RP-SVol$_{\mu}$ and RP-SVol$_{0}$ are almost indistinguishable, with Sharpe ratios just below that of the equally weighted portfolio.
The benchmark threshold $\tau_t=R^{I}_t$ brings the Sharpe ratios of the four strategies that adopt it within the range $0.872$--$0.896$.

\subsubsection{Nasdaq100}\label{subsubsec:nasdaq100}

\begin{table}[htbp]
\centering
\caption{Out-of-sample performance results for the Nasdaq100 dataset; units as in Table~\ref{tab:oos_dowjones}.}
\label{tab:oos_nasdaq}
\small
\renewcommand{\arraystretch}{1.15}
\begin{adjustbox}{width=\textwidth}
\begin{tabular}{lcccccccccccc}
\toprule
\textbf{Model} & $\mu^{\mathrm{ann}}$ & $\sigma^{\mathrm{ann}}$ & Sharpe & Sortino & Calmar & Omega & Rachev$_{10\%}$ & CVaR$_{5\%}$ & MaxDD & Ulcer & Turn & \#Assets \\
\midrule
EW & \cellcolor[RGB]{99,190,123}\textbf{22.81} & \cellcolor[RGB]{254,219,129}19.74 & \cellcolor[RGB]{99,190,123}\textbf{1.156} & \cellcolor[RGB]{99,190,123}\textbf{1.657} & \cellcolor[RGB]{99,190,123}\textbf{0.718} & \cellcolor[RGB]{99,190,123}\textbf{1.237} & \cellcolor[RGB]{254,219,129}0.971 & \cellcolor[RGB]{254,219,129}2.95 & \cellcolor[RGB]{158,207,126}31.78 & \cellcolor[RGB]{196,218,129}6.74 & \cellcolor[RGB]{138,201,125}5.98 & 84.00 \\
\midrule
MV & \cellcolor[RGB]{249,121,110}12.20 & \cellcolor[RGB]{118,196,124}14.04 & \cellcolor[RGB]{252,170,120}0.869 & \cellcolor[RGB]{252,170,120}1.207 & \cellcolor[RGB]{248,105,107}0.344 & \cellcolor[RGB]{252,186,123}1.177 & \cellcolor[RGB]{248,105,107}0.951 & \cellcolor[RGB]{128,198,125}2.04 & \cellcolor[RGB]{236,229,131}35.48 & \cellcolor[RGB]{158,207,126}6.49 & \cellcolor[RGB]{255,235,132}16.91 & 21.25 \\
MSV$_{0}$ & \cellcolor[RGB]{250,138,113}12.42 & \cellcolor[RGB]{138,201,125}14.12 & \cellcolor[RGB]{252,186,123}0.880 & \cellcolor[RGB]{252,186,123}1.222 & \cellcolor[RGB]{252,186,123}0.367 & \cellcolor[RGB]{252,170,120}1.175 & \cellcolor[RGB]{253,202,126}0.970 & \cellcolor[RGB]{128,198,125}2.04 & \cellcolor[RGB]{177,212,128}33.88 & \cellcolor[RGB]{216,224,130}6.89 & \cellcolor[RGB]{252,170,120}21.14 & 19.24 \\
MSV$_{\mu}$ & \cellcolor[RGB]{248,105,107}11.89 & \cellcolor[RGB]{99,190,123}\textbf{14.02} & \cellcolor[RGB]{251,154,116}0.848 & \cellcolor[RGB]{251,154,116}1.179 & \cellcolor[RGB]{250,138,113}0.348 & \cellcolor[RGB]{251,154,116}1.170 & \cellcolor[RGB]{252,186,123}0.968 & \cellcolor[RGB]{99,190,123}\textbf{2.02} & \cellcolor[RGB]{196,218,129}34.16 & \cellcolor[RGB]{236,229,131}6.93 & \cellcolor[RGB]{253,202,126}18.96 & 19.57 \\
MSV$_{R^{I}}$ & \cellcolor[RGB]{177,212,128}21.96 & \cellcolor[RGB]{252,186,123}21.62 & \cellcolor[RGB]{255,235,132}1.016 & \cellcolor[RGB]{255,235,132}1.455 & \cellcolor[RGB]{216,224,130}0.611 & \cellcolor[RGB]{226,227,130}1.205 & \cellcolor[RGB]{250,146,115}0.959 & \cellcolor[RGB]{252,178,121}3.24 & \cellcolor[RGB]{254,219,129}35.95 & \cellcolor[RGB]{252,170,120}9.57 & \cellcolor[RGB]{216,224,130}11.77 & 66.08 \\
\midrule
MaxSharpe & \cellcolor[RGB]{252,186,123}18.24 & \cellcolor[RGB]{250,138,113}22.12 & \cellcolor[RGB]{250,138,113}0.824 & \cellcolor[RGB]{250,138,113}1.168 & \cellcolor[RGB]{252,170,120}0.352 & \cellcolor[RGB]{248,113,109}1.166 & \cellcolor[RGB]{196,218,129}0.974 & \cellcolor[RGB]{250,138,113}3.40 & \cellcolor[RGB]{250,138,113}51.80 & \cellcolor[RGB]{250,138,113}22.97 & \cellcolor[RGB]{250,138,113}46.98 & 12.54 \\
MaxSortino$_{\mu}$ & \cellcolor[RGB]{216,224,130}20.14 & \cellcolor[RGB]{249,121,110}24.83 & \cellcolor[RGB]{249,121,110}0.811 & \cellcolor[RGB]{249,121,110}1.161 & \cellcolor[RGB]{251,154,116}0.349 & \cellcolor[RGB]{250,138,113}1.168 & \cellcolor[RGB]{118,196,124}0.993 & \cellcolor[RGB]{249,121,110}3.78 & \cellcolor[RGB]{249,121,110}57.73 & \cellcolor[RGB]{249,121,110}27.35 & \cellcolor[RGB]{249,121,110}50.04 & 10.62 \\
MaxSortino$_{0}$ & \cellcolor[RGB]{196,218,129}20.15 & \cellcolor[RGB]{248,105,107}25.11 & \cellcolor[RGB]{248,105,107}0.803 & \cellcolor[RGB]{248,105,107}1.150 & \cellcolor[RGB]{249,121,110}0.346 & \cellcolor[RGB]{248,113,109}1.166 & \cellcolor[RGB]{99,190,123}\textbf{0.996} & \cellcolor[RGB]{248,105,107}3.82 & \cellcolor[RGB]{248,105,107}58.25 & \cellcolor[RGB]{248,105,107}27.84 & \cellcolor[RGB]{248,105,107}50.58 & 10.32 \\
MaxSortino$_{R^{I}}$ & \cellcolor[RGB]{138,201,125}22.28 & \cellcolor[RGB]{251,154,116}21.83 & \cellcolor[RGB]{216,224,130}1.021 & \cellcolor[RGB]{226,227,130}1.460 & \cellcolor[RGB]{236,229,131}0.605 & \cellcolor[RGB]{255,227,130}1.204 & \cellcolor[RGB]{249,121,110}0.955 & \cellcolor[RGB]{251,154,116}3.28 & \cellcolor[RGB]{251,154,116}36.84 & \cellcolor[RGB]{251,154,116}9.86 & \cellcolor[RGB]{236,229,131}16.74 & 55.26 \\
\midrule
RP-Vol & \cellcolor[RGB]{236,229,131}19.53 & \cellcolor[RGB]{255,235,132}17.33 & \cellcolor[RGB]{118,196,124}1.127 & \cellcolor[RGB]{118,196,124}1.609 & \cellcolor[RGB]{138,201,125}0.623 & \cellcolor[RGB]{118,196,124}1.236 & \cellcolor[RGB]{245,232,131}0.972 & \cellcolor[RGB]{255,235,132}2.58 & \cellcolor[RGB]{138,201,125}31.36 & \cellcolor[RGB]{118,196,124}5.85 & \cellcolor[RGB]{118,196,124}5.93 & 84.00 \\
RP-SVol$_{\mu}$ & \cellcolor[RGB]{255,235,132}19.41 & \cellcolor[RGB]{236,229,131}17.27 & \cellcolor[RGB]{158,207,126}1.124 & \cellcolor[RGB]{158,207,126}1.604 & \cellcolor[RGB]{158,207,126}0.621 & \cellcolor[RGB]{148,204,126}1.235 & \cellcolor[RGB]{216,224,130}0.973 & \cellcolor[RGB]{226,227,130}2.57 & \cellcolor[RGB]{99,190,123}\textbf{31.25} & \cellcolor[RGB]{99,190,123}\textbf{5.80} & \cellcolor[RGB]{158,207,126}6.05 & 84.00 \\
RP-SVol$_{0}$ & \cellcolor[RGB]{254,219,129}19.38 & \cellcolor[RGB]{216,224,130}17.23 & \cellcolor[RGB]{138,201,125}1.125 & \cellcolor[RGB]{138,201,125}1.605 & \cellcolor[RGB]{177,212,128}0.620 & \cellcolor[RGB]{148,204,126}1.235 & \cellcolor[RGB]{245,232,131}0.972 & \cellcolor[RGB]{226,227,130}2.57 & \cellcolor[RGB]{118,196,124}31.29 & \cellcolor[RGB]{138,201,125}5.86 & \cellcolor[RGB]{99,190,123}\textbf{5.71} & 84.00 \\
RP-SVol$_{R^{I}}$ & \cellcolor[RGB]{118,196,124}22.62 & \cellcolor[RGB]{253,202,126}21.49 & \cellcolor[RGB]{196,218,129}1.052 & \cellcolor[RGB]{196,218,129}1.509 & \cellcolor[RGB]{118,196,124}0.651 & \cellcolor[RGB]{196,218,129}1.213 & \cellcolor[RGB]{252,170,120}0.960 & \cellcolor[RGB]{253,202,126}3.22 & \cellcolor[RGB]{216,224,130}34.76 & \cellcolor[RGB]{253,202,126}9.18 & \cellcolor[RGB]{177,212,128}6.89 & 84.00 \\
\midrule
MDP-Vol & \cellcolor[RGB]{253,202,126}18.78 & \cellcolor[RGB]{196,218,129}16.94 & \cellcolor[RGB]{177,212,128}1.109 & \cellcolor[RGB]{177,212,128}1.575 & \cellcolor[RGB]{255,235,132}0.520 & \cellcolor[RGB]{177,212,128}1.227 & \cellcolor[RGB]{138,201,125}0.992 & \cellcolor[RGB]{196,218,129}2.49 & \cellcolor[RGB]{253,202,126}36.13 & \cellcolor[RGB]{254,219,129}7.64 & \cellcolor[RGB]{254,219,129}18.84 & 24.97 \\
MDP-SVol$_{\mu}$ & \cellcolor[RGB]{251,154,116}16.49 & \cellcolor[RGB]{177,212,128}16.54 & \cellcolor[RGB]{253,202,126}0.997 & \cellcolor[RGB]{253,202,126}1.412 & \cellcolor[RGB]{253,202,126}0.451 & \cellcolor[RGB]{253,202,126}1.202 & \cellcolor[RGB]{167,210,127}0.986 & \cellcolor[RGB]{167,210,127}2.40 & \cellcolor[RGB]{252,186,123}36.59 & \cellcolor[RGB]{177,212,128}6.60 & \cellcolor[RGB]{251,154,116}21.68 & 23.35 \\
MDP-SVol$_{0}$ & \cellcolor[RGB]{252,170,120}16.58 & \cellcolor[RGB]{158,207,126}16.51 & \cellcolor[RGB]{254,219,129}1.004 & \cellcolor[RGB]{254,219,129}1.419 & \cellcolor[RGB]{254,219,129}0.453 & \cellcolor[RGB]{255,227,130}1.204 & \cellcolor[RGB]{167,210,127}0.986 & \cellcolor[RGB]{167,210,127}2.40 & \cellcolor[RGB]{252,170,120}36.62 & \cellcolor[RGB]{255,235,132}7.03 & \cellcolor[RGB]{252,186,123}21.10 & 22.77 \\
MDP-SVol$_{R^{I}}$ & \cellcolor[RGB]{158,207,126}22.04 & \cellcolor[RGB]{252,170,120}21.63 & \cellcolor[RGB]{236,229,131}1.019 & \cellcolor[RGB]{226,227,130}1.460 & \cellcolor[RGB]{196,218,129}0.615 & \cellcolor[RGB]{226,227,130}1.205 & \cellcolor[RGB]{250,146,115}0.959 & \cellcolor[RGB]{252,178,121}3.24 & \cellcolor[RGB]{255,235,132}35.84 & \cellcolor[RGB]{252,186,123}9.56 & \cellcolor[RGB]{196,218,129}11.72 & 66.58 \\
\midrule
Market index & 20.56 & 21.96 & 0.936 & 1.336 & 0.587 & 1.186 & 0.952 & 3.29 & 35.04 & 9.64 & --- & --- \\
\bottomrule
\end{tabular}
\end{adjustbox}
\end{table}

In Table~\ref{tab:oos_nasdaq}, we observe that the equally weighted portfolio has the highest annualized return and the best Sharpe ($1.156$), Sortino, Calmar and Omega ratios.
Also in this case, MV, MSV$_{0}$ and MSV$_{\mu}$ show the lowest volatility and CVaR$_{5\%}$ with the worst annualized returns, and Sharpe ratios below that of the market index ($0.936$).
MaxSharpe, MaxSortino$_{\mu}$ and MaxSortino$_{0}$ present the lowest Sharpe ratios of the comparison, together with the worst values of all risk measures and turnover.
The RP-Vol, RP-SVol$_{\mu}$ and RP-SVol$_{0}$ portfolios are again almost indistinguishable, with Sharpe ratios second only to that of the equally weighted portfolio and the lowest Maximum Drawdown and Ulcer index.
In the most-diversified family, MDP-Vol attains a higher Sharpe ratio ($1.109$) than MDP-SVol$_{\mu}$ ($0.997$) and MDP-SVol$_{0}$ ($1.004$), a difference that is not significant in Table~\ref{tab:tests}.
The benchmark threshold brings the Sharpe ratios of the four strategies that adopt it within the range $1.016$--$1.052$, all above that of the market index.

\subsubsection{Comparison across the two datasets}\label{subsubsec:all}

Table~\ref{tab:tests} reports the difference between the annualized out-of-sample Sharpe ratios, net of transaction costs, of strategies of the same model class that differ in the risk measure (Panel~A) or in the threshold (Panels~B and~C).
The $95\%$ confidence intervals are obtained with a studentized circular block bootstrap \citep{politis1992circular,ledoit2008robust}, with blocks of $21$ days, $1999$ replications and Bartlett HAC bandwidth $9$ on the DowJones and $8$ on the Nasdaq100.
The $p$-values are adjusted with the procedure of \citet{holm1979simple} within each panel and dataset, and the inference is conditional on the out-of-sample return series produced by the backtest.
The tests against the equally weighted portfolio, the MV model and the market index are reported in the Supplementary Material.

\begin{table}[htbp]
\centering
\caption{Within-family differences of the out-of-sample Sharpe ratio, in natural units: point estimate, unadjusted equal-tailed $95\%$ bootstrap confidence interval and Holm-adjusted $p$-value. With the $1/N^{\mathrm{out}}$ variance estimator, the differences may deviate by $0.0001$ from Tables~\ref{tab:oos_dowjones}--\ref{tab:oos_nasdaq}.}
\label{tab:tests}
\small
\renewcommand{\arraystretch}{1.15}
\begin{adjustbox}{max width=\textwidth}
\begin{tabular}{lcccc}
\toprule
& \multicolumn{2}{c}{DowJones} & \multicolumn{2}{c}{Nasdaq100} \\
\cmidrule(lr){2-3}\cmidrule(lr){4-5}
& $\Delta$ [95\% CI] & $p_{\mathrm{Holm}}$ & $\Delta$ [95\% CI] & $p_{\mathrm{Holm}}$ \\
\midrule
\multicolumn{5}{l}{\textit{Panel A. Risk measure: downside minus variance-based model, $\tau=\mu^\top x$}} \\
MSV$_{\mu}$ $-$ MV & $-$0.0077 [$-$0.0575, 0.0427] & 1.000 & $-$0.0209 [$-$0.1145, 0.0718] & 1.000 \\
MaxSortino$_{\mu}$ $-$ MaxSharpe & $-$0.0247 [$-$0.0880, 0.0402] & 1.000 & $-$0.0133 [$-$0.1382, 0.1030] & 1.000 \\
RP-SVol$_{\mu}$ $-$ RP-Vol & $-$0.0010 [$-$0.0075, 0.0054] & 1.000 & $-$0.0033 [$-$0.0125, 0.0062] & 1.000 \\
MDP-SVol$_{\mu}$ $-$ MDP-Vol & 0.0429 [$-$0.0325, 0.1222] & 1.000 & $-$0.1120 [$-$0.2372, 0.0070] & 0.292 \\
\midrule
\multicolumn{5}{l}{\textit{Panel B. Threshold: $\tau=0$ minus $\tau=\mu^\top x$ (e.g.\ MSV$_{0}$ $-$ MSV$_{\mu}$)}} \\
MSV & 0.0208 [$-$0.0288, 0.0695] & 0.836 & 0.0318 [$-$0.0401, 0.1063] & 1.000 \\
MaxSortino & $-$0.0051 [$-$0.0141, 0.0040] & 0.836 & $-$0.0082 [$-$0.0239, 0.0078] & 1.000 \\
RP-SVol & 0.0015 [$-$0.0046, 0.0077] & 0.836 & 0.0012 [$-$0.0091, 0.0112] & 1.000 \\
MDP-SVol & 0.0183 [$-$0.0110, 0.0475] & 0.836 & 0.0074 [$-$0.0371, 0.0539] & 1.000 \\
\midrule
\multicolumn{5}{l}{\textit{Panel C. Threshold: $\tau_t=R^{I}_t$ minus $\tau=\mu^\top x$ (e.g.\ MSV$_{R^{I}}$ $-$ MSV$_{\mu}$)}} \\
MSV & 0.1055 [$-$0.1995, 0.4148] & 0.979 & 0.1681 [$-$0.3413, 0.7438] & 1.000 \\
MaxSortino & 0.0352 [$-$0.3454, 0.3961] & 0.979 & 0.2096 [$-$0.2049, 0.6589] & 1.000 \\
RP-SVol & $-$0.1481 [$-$0.2374, $-$0.0661] & 0.006 & $-$0.0714 [$-$0.3275, 0.2075] & 1.000 \\
MDP-SVol & $-$0.1989 [$-$0.3965, $-$0.0065] & 0.137 & 0.0224 [$-$0.3852, 0.4631] & 1.000 \\
\bottomrule
\end{tabular}
\end{adjustbox}
\end{table}

In Panel~A, we observe that at the below-mean threshold no downside strategy differs significantly from its variance-based counterpart, in any model class and on either dataset.
The confidence intervals are narrow only for risk parity, where the difference between RP-SVol$_{\mu}$ and RP-Vol is $-0.0010$ on the DowJones and $-0.0033$ on the Nasdaq100.
As a margin for equivalence we take the annual transaction costs of RP-Vol expressed in Sharpe-ratio units, $\delta = 12\,c_{\mathrm{tc}}\,\Turn/\sigma^{\mathrm{ann}}$, with $\Turn$, as a fraction, and $\sigma^{\mathrm{ann}}$ its turnover and annualized volatility.
With these margins ($0.018$ on the DowJones and $0.021$ on the Nasdaq100), the two one-sided tests procedure of \citet{schuirmann1987comparison} declares RP-Vol and RP-SVol$_{\mu}$ equivalent on both datasets.
This rule was chosen after the analysis and not pre-specified, so the equivalence is exploratory, and its sensitivity to the bootstrap settings is reported in Supplementary Table~S.5.
In the minimum-risk, reward-to-risk and most-diversified classes the confidence intervals are instead too wide either to separate the two risk measures or to establish their equivalence.

Panel~B shows that the zero and the below-mean thresholds lead to Sharpe ratios that are not statistically different in any model class, on either dataset.
For the Sortino and risk-parity models the two thresholds behave almost identically, consistently with the coincidence condition of Section~\ref{subsec:sortino}, since at the daily frequency $\mu^\top x$ is small relative to the semideviation.
In the minimum-risk and most-diversified classes the point estimates favor $\tau=0$ on both datasets, but the differences are not significant.

The benchmark threshold $\tau_t=R^{I}_t$ has the largest effect (Panel~C), and the four strategies that adopt it deliver nearly the same return, volatility and Sharpe ratio, although not the same turnover.
We interpret this convergence as an effect of the common benchmark, since the below-mean versions do not converge.
Our design, however, does not separate the benchmark from its one-sided treatment, and for MaxSortino$_{R^{I}}$ the numerator also becomes the expected active return.
After the Holm adjustment, the only significant difference of Table~\ref{tab:tests} concerns risk parity on the DowJones, where the benchmark threshold lowers the Sharpe ratio by $0.148$.\footnote{The unadjusted $p$-value is $0.0015$; adjusting over the eight comparisons of Panel~C on the two datasets gives $0.012$ under both the procedure of \citet{holm1979simple} and that of \citet{benjamini1995controlling}, so that the difference remains significant at the $5\%$ level over this larger family.}
On the Nasdaq100 the reduction is $0.071$ and not significant, and in the other classes the differences are of mixed sign and not significant after the adjustment.
In risk parity, where both effects are estimated most precisely, the point estimates of the benchmark-threshold effect are far larger than those of the replacement of volatility with semivolatility.
Comparisons of downside-risk models should thus state the threshold convention that they adopt.

\section{Conclusions and further research}\label{sec:conclusion}

In this paper, we have studied downside risk in long-only portfolio selection beyond minimum-risk models, where semivariance and semivolatility yield the same portfolio.
We have classified the risk-budgeting, most-diversified and reward-to-risk models by the degree of positive homogeneity of the risk measure, since degree one keeps the diversification and reward-to-risk ratios scale invariant and makes the risk contributions add up to the portfolio risk.
We have made the threshold explicit through a non-degeneracy condition, stated for the portfolio mean, zero and a market-index return, under which the risk-parity, most-diversified and Sortino-ratio problems are well posed.
Under this condition, the two ratio problems reduce to convex programs by a known reduction \citep{schaible1974parameter,stoyanov2007optimal,cesarone2025return}, and the differentiability of semivolatility, a special case of a result of \citet{fischer2003risk}, ensures that the risk-budgeting program attains the prescribed budgets.
We have also characterized when below-mean semivolatility is additive.

Out of sample, we have compared seventeen long-only strategies on the DowJones and Nasdaq100 datasets of total returns and tested the Sharpe-ratio differences within each family by block bootstrap.
Replacing volatility with below-mean semivolatility produces no significant difference in the Sharpe ratio, and in risk parity the volatility and semivolatility portfolios are equivalent, in an exploratory sense.
On the DowJones, MDP-SVol$_{0}$ and MDP-SVol$_{\mu}$ attain the highest and second-highest Sharpe ratios, but the difference between MDP-SVol$_{\mu}$ and MDP-Vol is not significant and changes sign on the Nasdaq100, where the equally weighted portfolio ranks first.
The zero and below-mean thresholds give Sharpe ratios that are not statistically different, whereas the market-index threshold changes them by mostly larger but often imprecise amounts.
The only significant change is a reduction of $0.148$ for risk parity on the DowJones, which survives the Holm adjustment also over the market-index comparisons of both datasets.
Comparisons of downside-risk models should therefore state the threshold convention they adopt.

These findings rest on datasets of current index constituents, a selection that induces a survivorship and look-ahead bias in favor of every strategy relative to the market index, so that the results of the strategies using its return as threshold should be read with particular caution.
Further research might extend the taxonomy to higher-order lower partial moments (see, e.g., \citealt{ogryczak2001consistency,cesarone2025semivariance}) and repeat the comparison on point-in-time index constituents.
\appendix

\section{Additivity of semivolatility}\label{app:additivity}

This appendix proves the characterization announced in Remark~\ref{rem:additivity} and derives the closed-form risk-parity portfolio it yields.

\begin{lemma}[Additivity of semivolatility]\label{lem:sv-additivity-pair}
Let $X,Y\in L^2$.
Then
\begin{equation}\label{eq:sv-add-pair}
\SemiVol_\mu(X+Y) = \SemiVol_\mu(X) + \SemiVol_\mu(Y)
\end{equation}
if and only if the downside deviations are \emph{positively collinear}: there exist $\alpha,\beta\ge 0$, not both zero, such that $\alpha\,\bigl(X-\E[X]\bigr)_- = \beta\,\bigl(Y-\E[Y]\bigr)_-$ $\mathbb{P}$-a.s.
In this case $\bigl(X-\E[X]\bigr)\bigl(Y-\E[Y]\bigr) \ge 0$ $\mathbb{P}$-a.s.
\end{lemma}

\begin{proof}
Set $\xi_X := X-\E[X]$ and $\xi_Y := Y-\E[Y]$, so that $\SemiVol_\mu(X+Y) = \|(\xi_X+\xi_Y)_-\|_{L^2}$.
Two elementary facts are used.
\begin{enumerate}[label=(\Roman*)]
    \item For $a,b\in\R$, $(a+b)_- \le a_- + b_-$, with equality iff
    $ab\ge 0$ (case analysis on signs).
    \item For $U,V\in L^2_+$, the triangle inequality
    $\|U+V\|_{L^2} \le \|U\|_{L^2}+\|V\|_{L^2}$ is tight iff
    $\langle U,V\rangle = \|U\|_{L^2}\|V\|_{L^2}$, i.e.\ iff there exist
    $\alpha,\beta\ge0$, not both zero, with $\alpha U = \beta V$
    $\mathbb{P}$-a.s.\ (equality case of Cauchy--Schwarz). The symmetric form
    is needed: the one-sided form $V=\lambda U$ fails when $U=0\neq V$, a case
    in which the inequality is tight.
\end{enumerate}
Applying~(I) pointwise and~(II) to $(\xi_X)_-$ and $(\xi_Y)_-$,
\[
\SemiVol_\mu(X+Y) \le \|(\xi_X)_- + (\xi_Y)_-\|_{L^2} \le \|(\xi_X)_-\|_{L^2}+\|(\xi_Y)_-\|_{L^2} = \SemiVol_\mu(X)+\SemiVol_\mu(Y).
\]
Since $0\le(\xi_X+\xi_Y)_-\le(\xi_X)_-+(\xi_Y)_-$ pointwise, equality in the first inequality forces the nonnegative random variable $((\xi_X)_-+(\xi_Y)_-)^2-(\xi_X+\xi_Y)_-^2$ to have zero expectation, hence to vanish a.s., which by~(I) means $\xi_X\xi_Y\ge0$ a.s.; equality in the second holds iff the collinearity condition holds, by~(II).
Hence~\eqref{eq:sv-add-pair} holds iff $\xi_X\xi_Y\ge0$ a.s.\ and the collinearity condition holds.
It remains to show that collinearity implies $\xi_X\xi_Y\ge0$ a.s.
If $\alpha,\beta>0$, then $\{\xi_X<0\}=\{\xi_Y<0\}$ a.s., so $\xi_X$ and $\xi_Y$ never have opposite signs.
If $\alpha=0$, then $\beta>0$ and $(\xi_Y)_-=0$ a.s., so $\xi_Y\ge0$ with $\E[\xi_Y]=0$, whence $\xi_Y=0$ a.s.; the case $\beta=0$ is symmetric.
\end{proof}

The sign condition is the MAD-additivity condition of Remark~\ref{rem:additivity}, and both comparisons stated there are strict: on four equiprobable states, $X=(-1,-1,1,1)$ is MAD-additive but not $\SemiVol_\mu$-additive with $Y=(-1,-2,\tfrac32,\tfrac32)$, and $\SemiVol_\mu$-additive but not $\sd$-additive with $Y=(-1,-1,\tfrac12,\tfrac32)$.
The next result is the semivolatility analogue of \citet[Theorem~6]{ararat2021mad}.

\begin{theorem}[Multivariate additivity of semivolatility]
\label{thm:sv-additivity-multi}
Let $Y_1,\dots,Y_n\in L^2$.
The following are equivalent:
\begin{enumerate}[label=(\roman*)]
    \item $\displaystyle \SemiVol_\mu\!\left(\sum_{i=1}^n x_i Y_i\right)
    = \sum_{i=1}^n x_i\,\SemiVol_\mu(Y_i)$
    for every $x\in\R^n_+$;
    \item the identity in~(i) holds for every $x\in\Delta$;
    \item $\displaystyle \SemiVol_\mu\!\left(\sum_{i\in I} Y_i\right)
    = \sum_{i\in I} \SemiVol_\mu(Y_i)$
    for every nonempty $I\subseteq\{1,\dots,n\}$;
    \item for every $i,j\in\{1,\dots,n\}$ with $i\ne j$, there exist
    $\alpha_{ij},\beta_{ij}\ge 0$, not both zero, such that
    \[
    \alpha_{ij}\,\bigl(Y_i-\E[Y_i]\bigr)_-
    = \beta_{ij}\,\bigl(Y_j-\E[Y_j]\bigr)_-\
    \mathbb{P}\text{-a.s.};
    \]
    \item (\emph{common downside factor}) there exist $V\in L^2_+$ and
    nonnegative constants $\kappa_1,\dots,\kappa_n$ such that
    \[
    \bigl(Y_i-\E[Y_i]\bigr)_- = \kappa_i\,V \quad \mathbb{P}\text{-a.s.}
    \quad\text{for every } i\in\{1,\dots,n\}.
    \]
\end{enumerate}
\end{theorem}

\begin{proof}
Set $\xi_i := Y_i - \E[Y_i]$ for $i=1,\dots,n$.
We prove (i) $\Rightarrow$ (ii), (iii), then (ii) $\Rightarrow$ (i), and finally (iii) $\Rightarrow$ (iv) $\Rightarrow$ (v) $\Rightarrow$ (i).

\emph{(i) $\Rightarrow$ (ii), (iii).} Immediate, by specialization of~$x$.

\emph{(ii) $\Rightarrow$ (i).} Let $x\in\R^n_+$.
The case $x=0$ is trivial.
If $x\neq 0$, set $a := \sum_{i=1}^n x_i > 0$; then $x/a \in \Delta$ and, by degree-one homogeneity of $\SemiVol_\mu$,
\[
\SemiVol_\mu\!\left(\sum_{i=1}^n x_i Y_i\right) = a\,\SemiVol_\mu\!\left(\sum_{i=1}^n \tfrac{x_i}{a} Y_i\right) \stackrel{\text{(ii)}}{=} a \sum_{i=1}^n \tfrac{x_i}{a}\,\SemiVol_\mu(Y_i) = \sum_{i=1}^n x_i \,\SemiVol_\mu(Y_i).
\]

\emph{(iii) $\Rightarrow$ (iv).} Fix $i\ne j$ and apply~(iii) with $I=\{i,j\}$: $\SemiVol_\mu(Y_i+Y_j) = \SemiVol_\mu(Y_i)+\SemiVol_\mu(Y_j)$.
Lemma~\ref{lem:sv-additivity-pair} then yields~(iv) for the pair $(Y_i,Y_j)$.

\emph{(iv) $\Rightarrow$ (v).} If $(\xi_i)_- = 0$ $\mathbb{P}$-a.s.\ for every $i$, take $V:=0$ and $\kappa_i:=0$.
Otherwise, fix an index $\ell$ such that $(\xi_\ell)_-$ is not $\mathbb{P}$-a.s.\ zero, set $V := (\xi_\ell)_-$ and $\kappa_\ell := 1$.
Fix $i\neq\ell$ and let $\alpha_{\ell i},\beta_{\ell i}\ge 0$, not both zero, be the constants provided by~(iv), so that $\alpha_{\ell i}\,(\xi_\ell)_- = \beta_{\ell i}\,(\xi_i)_-$ $\mathbb{P}$-a.s.
If $\beta_{\ell i}=0$, then $\alpha_{\ell i}>0$ and $(\xi_\ell)_-=0$ $\mathbb{P}$-a.s., contradicting the choice of $\ell$; hence $\beta_{\ell i}>0$, and $\kappa_i := \alpha_{\ell i}/\beta_{\ell i}$ gives $(\xi_i)_- = \kappa_i\,V$ $\mathbb{P}$-a.s.

\emph{(v) $\Rightarrow$ (i).} Let $x\in\R^n_+$, $\xi := \sum_i x_i\,\xi_i$ and $J:=\{i:\kappa_i>0\}$.
For $i\notin J$, $(\xi_i)_-=0$ a.s.\ and $\E[\xi_i]=0$ give $\xi_i=0$ a.s.
For $i\in J$, $\{\xi_i<0\}=\{V>0\}$ a.s., and $\xi_i=-\kappa_iV$ on $\{V>0\}$.
Hence $\xi=-\bigl(\sum_i x_i\kappa_i\bigr)V$ on $\{V>0\}$ and $\xi\ge0$ on $\{V=0\}$, so that $\xi_- = \bigl(\sum_i x_i \kappa_i\bigr) V$ $\mathbb{P}$-a.s.
Taking the $L^2$-norm and using $\SemiVol_\mu(Y_i) = \kappa_i\|V\|_{L^2}$,
\[
\SemiVol_\mu\!\Bigl(\sum_i x_i Y_i\Bigr) = \Bigl(\sum_i x_i \kappa_i\Bigr)\|V\|_{L^2} = \sum_i x_i\,\SemiVol_\mu(Y_i).
\]
\end{proof}

Under this condition, the risk-parity problem~\eqref{prob:rp-semivol} admits an explicit solution, the semivolatility counterpart of the closed-form MAD-RP portfolio of \citet[Proposition~13]{ararat2021mad}.

\begin{proposition}[Closed-form SV-RP portfolio]\label{prop:svrp-closed-form}
Suppose that Assumption~\ref{ass:nondeg} holds and that the asset returns $R_1,\dots,R_n$ satisfy condition~(iv) of Theorem~\ref{thm:sv-additivity-multi}.
Let $b_1,\dots,b_n>0$ with $\sum_{i=1}^n b_i=1$.
Then the SV-RP portfolio $x^{\mathrm{RP}}=x^\star$ of Proposition~\ref{prop:rp-budget} is given in closed form by
\begin{equation}\label{eq:svrp-closedform}
x_i^{\mathrm{RP}}
=\frac{b_i\,\SemiVol_\mu(R_i)^{-1}}%
      {\sum_{j=1}^n b_j\,\SemiVol_\mu(R_j)^{-1}},
\qquad i=1,\dots,n.
\end{equation}
For $b_i=1/n$ the weights are proportional to $\SemiVol_\mu(R_i)^{-1}$.
\end{proposition}

\begin{proof}
By Proposition~\ref{prop:nondeg-equiv}, Assumption~\ref{ass:nondeg} implies $\SemiVol_\mu(R_i)>0$ for every~$i$, so the expression in \eqref{eq:svrp-closedform} is well defined.
Under condition~(iv), Theorem~\ref{thm:sv-additivity-multi} gives $\SemiVolP(x)=\sum_{i=1}^n x_i\,\SemiVol_\mu(R_i)$ for $x\in\R^n_+$, so $\SemiVolP$ is linear on $\R^n_+$ and the downside risk contributions on $\R^n_{++}$ are $\RC_i(x)=x_i\,\SemiVol_\mu(R_i)$.
Since $y^\star\in\R^n_{++}$, also $x^{\mathrm{RP}}\in\R^n_{++}$, and Proposition~\ref{prop:rp-budget} gives $x_i^{\mathrm{RP}}\,\SemiVol_\mu(R_i)=b_i\,\SemiVolP(x^{\mathrm{RP}})$ for every~$i$.
Summing $x_i^{\mathrm{RP}}=b_i\,\SemiVolP(x^{\mathrm{RP}})/\SemiVol_\mu(R_i)$ over~$i$ and using $\sum_{i=1}^n x_i^{\mathrm{RP}}=1$ gives $\SemiVolP(x^{\mathrm{RP}})=\bigl(\sum_{i=1}^n b_i\,\SemiVol_\mu(R_i)^{-1}\bigr)^{-1}$, and substituting back yields~\eqref{eq:svrp-closedform}.
\end{proof}

\section{Proofs}\label{app:proofs}

\subsection*{Proof of Proposition~\ref{prop:nondeg-equiv}}

\begin{proof}
(i) $\Leftrightarrow$ (ii).
Set $Z := (R-\mu)^\top x$.
Since $\E[R-\mu] = 0$, we have $\E[Z] = 0$.
If $Z_- = 0$ $\mathbb{P}$-a.s., then $Z \ge 0$ $\mathbb{P}$-a.s.; combined with $\E[Z] = 0$, this forces $Z = 0$ $\mathbb{P}$-a.s.
Conversely, $Z = 0$ $\mathbb{P}$-a.s.\ trivially implies $Z_- = 0$ $\mathbb{P}$-a.s.
Hence, $(R_P(x)-\mu^\top x)_- = Z_- = 0$ $\mathbb{P}$-a.s.\ if and only if $(R-\mu)^\top x = 0$ $\mathbb{P}$-a.s., which gives the equivalence of (i) and (ii).

(i) $\Leftrightarrow$ (iii).
By definition, $\SemiVol_\mu(R_P(x))^2 = \E\!\bigl[\bigl((R-\mu)^\top x\bigr)_-^2\bigr]$, which vanishes if and only if $\bigl((R-\mu)^\top x\bigr)_- = 0$ $\mathbb{P}$-a.s.

(ii) $\Leftrightarrow$ (iv).
Since $\sd(R_P(x))^2 = \E\!\bigl[\bigl((R-\mu)^\top x\bigr)^2\bigr]$, the variance vanishes if and only if $(R-\mu)^\top x = 0$ $\mathbb{P}$-a.s.
\end{proof}

\subsection*{Proof of Lemma~\ref{lem:sv-differentiable}}

\begin{proof}
Write $Z(x):=(R-\mu)^\top x$ and $\varphi(x):=\SemiVolP(x)^2=\E[Z(x)_-^2]$, so that $\SemiVolP=\sqrt{\varphi}$.
Openness of $\mathcal{W}$ follows from the continuity of $\SemiVolP$, which is a consequence of its convexity and finiteness on $\R^n$.

The scalar function $u\mapsto u_-^2$ is continuously differentiable on $\R$ with derivative $-2u_-$ (both one-sided derivatives vanish at $u=0$); this, and no continuity of the distribution of $R$, is what the argument uses, whereas $u\mapsto u_-$ is not differentiable at the origin.

Fix $x$ and a direction $e_i$.
For $|h|\le 1$ the difference quotients of $h\mapsto Z(x+he_i)_-^2$ are bounded, by the mean value theorem, by $2\,|R_i-\mu_i|\,\bigl(|Z(x)|+|R_i-\mu_i|\bigr)\in L^1$, since $R\in (L^2)^n$.
Dominated convergence therefore allows differentiation under the expectation, giving
\[
\frac{\partial \varphi(x)}{\partial x_i} =-2\,\E\!\left[(R_i-\mu_i)\,Z(x)_-\right],
\]
and these partial derivatives are continuous in $x$, since $|Z(x)_--Z(x')_-|\le|Z(x)-Z(x')|$ and the Cauchy--Schwarz inequality give $\bigl|\E[(R_i-\mu_i)(Z(x)_--Z(x')_-)]\bigr|\le \|R_i-\mu_i\|_{L^2}\,\|(R-\mu)^\top(x-x')\|_{L^2}$; hence $\varphi\in C^1(\R^n)$, with Lipschitz gradient.
On $\mathcal{W}$ we have $\varphi>0$, so $\SemiVolP=\sqrt{\varphi}$ is $C^1$ there by the chain rule, with $\partial_i\SemiVolP=\partial_i\varphi/(2\sqrt{\varphi})$, which is~\eqref{eq:sv-gradient}.

Finally, under Assumption~\ref{ass:nondeg}, Proposition~\ref{prop:nondeg-equiv}(iii) gives $\SemiVolP(x)>0$ for every $x\in\R^n_+\setminus\{0\}$, i.e.\ $\R^n_+\setminus\{0\}\subseteq\mathcal{W}$.
\end{proof}

\subsection*{Existence and uniqueness for Problem~\eqref{prob:rp-semivol}}

The argument specializes \citet[Theorems~1 and~2]{cetingoz2024risk} to below-mean semivolatility, positivity on $\R^n_+\setminus\{0\}$ being supplied by Assumption~\ref{ass:nondeg}; we report it for completeness.

\begin{proof}
Let $\psi(y):=\sum_{i=1}^n b_i\log(y_i)$ and $F:=\{y\in\R^n_{++}:\psi(y)\ge\gamma\}$, the feasible set of Problem~\eqref{prob:rp-semivol}.
The set $F$ is nonempty ($e^{\gamma}\one\in F$) and closed in $\R^n$: if $y^k\in F$, $y^k\to\bar y$ and $\bar y_i=0$ for some $i$, then $\psi(y^k)\to-\infty$.
Under Assumption~\ref{ass:nondeg}, $m:=\min\{\SemiVolP(y):y\in\R^n_+,\ \|y\|=1\}>0$ and, by degree-one homogeneity, $\SemiVolP(y)\ge m\|y\|$ on $\R^n_+$, so the sublevel sets of $\SemiVolP$ on $F$ are compact and an optimal solution exists.
Let $y^1\neq y^2$ both be optimal with common value $v=\SemiVolP(y^1)>0$, and put $\bar y:=\tfrac12(y^1+y^2)$.
Since $b_i>0$ for every $i$, $\psi$ is strictly concave on $\R^n_{++}$, whence $\psi(\bar y)>\tfrac12\psi(y^1)+\tfrac12\psi(y^2)\ge \gamma$, namely the logarithmic constraint is slack at $\bar y$.
Convexity of $\SemiVolP$ gives $\SemiVolP(\bar y)\le v$ and, by continuity of $\psi$, there is $\alpha\in(0,1)$ with $\psi(\alpha\bar y)\ge \gamma$, so that $\alpha\bar y$ is feasible.
Degree-one homogeneity then yields $\SemiVolP(\alpha\bar y)=\alpha\,\SemiVolP(\bar y)\le\alpha v<v$, which contradicts optimality.
\end{proof}

\subsection*{Proof of Proposition~\ref{prop:rp-budget}}

\begin{proof}
Problem~\eqref{prob:rp-semivol} is convex and satisfies Slater's condition (take $y_i \equiv e^{\gamma+1}$), so the KKT conditions characterize optimality; under Assumption~\ref{ass:nondeg}, Lemma~\ref{lem:sv-differentiable} guarantees that $\SemiVolP$ is continuously differentiable at every feasible point.
Stationarity yields $\partial_i \SemiVolP(y^\star) = \lambda\,b_i/y_i^\star$ for some $\lambda\ge0$, and multiplying by $y_i^\star$ and summing, Euler's identity for degree-one functionals gives $\SemiVolP(y^\star) = \lambda\sum_i b_i = \lambda>0$, so that the logarithmic constraint is active by complementary slackness and
\[
y_i^\star\,\frac{\partial \SemiVolP(y^\star)}{\partial y_i} = b_i\,\SemiVolP(y^\star), \qquad i=1,\dots,n.
\]

Setting $x^\star:=y^\star/\one^\top y^\star$, degree-zero homogeneity of $\nabla \SemiVolP$ (differentiate $\SemiVolP(ty)=t\,\SemiVolP(y)$ in $y$) gives $\partial_i \SemiVolP(x^\star) = \partial_i \SemiVolP(y^\star)$, and degree-one homogeneity of $\SemiVolP$ gives $\SemiVolP(x^\star) = \SemiVolP(y^\star)/\one^\top y^\star$.
Substituting,
\[
\RC_i(x^\star) = x_i^\star\,\frac{\partial \SemiVolP(x^\star)}{\partial x_i} = \frac{y_i^\star\,\partial_i \SemiVolP(y^\star)}{\one^\top y^\star} = \frac{b_i\,\SemiVolP(y^\star)}{\one^\top y^\star} = b_i\,\SemiVolP(x^\star).
\]
Since the feasible set for $\gamma+c$ is $e^{c}$ times that for $\gamma$, the unique optimal solution scales by $e^{c}$, so that $x^\star$ does not depend on $\gamma$.
Conversely, if $x\in\Delta$ satisfies $\RC_i(x)=b_i\,\SemiVolP(x)$ for every $i$, then $x\in\R^n_{++}$, since $\SemiVolP(x)>0$ and $b_i>0$, and $x$ satisfies the KKT conditions with $\lambda=\SemiVolP(x)$ and $\gamma=\sum_i b_i\log(x_i)$, so that $x=x^\star$.
\end{proof}

\subsection*{Proof of Lemma~\ref{lem:schaible}}

\begin{proof}
Set $F:=\{y\in C:\ c^\top y=1\}$, which is closed and convex.

\emph{Feasibility.} Let $\bar x\in\Delta_C^+$.
Since $C$ is a cone and $c^\top\bar x>0$, the point $\bar y:=\bar x/(c^\top\bar x)$ belongs to $C$ and satisfies $c^\top\bar y=1$, so $F\neq\emptyset$.

\emph{Attainment.} The set $K:=\{y\in C:\|y\|=1\}$ is compact, and nonempty because $\Delta_C^+\neq\emptyset$ implies $C\neq\{0\}$; $\rho$ is continuous and strictly positive on $K$, so that $m:=\min_{K}\rho>0$.
By degree-one homogeneity, $\rho(y)\ge m\|y\|$ for every $y\in C$, hence the sublevel sets of $\rho$ relative to $F$ are bounded, and $\rho$ attains its minimum on $F$ by the Weierstrass theorem.
Note also that every $y\in F$ is nonzero, so $\rho(y)>0$ on $F$.

\emph{Equivalence.} Define $\Phi(x):=x/(c^\top x)$ for $x\in\Delta_C^+$ and $\Psi(y):=y/(\one^\top y)$ for $y\in F$.
For $x\in\Delta_C^+$ we have $\Phi(x)\in F$ and, by degree-one homogeneity,
\[
\frac{c^\top x}{\rho(x)}=\frac{1}{\rho\bigl(\Phi(x)\bigr)} .
\]
Conversely, let $y\in F$.
Since $y\in C\subseteq\R^n_+$ and $y\neq0$, we have $\one^\top y>0$, so $\Psi(y)\in C\cap\Delta$ and $c^\top\Psi(y)=1/(\one^\top y)>0$, namely $\Psi(y)\in\Delta_C^+$, and
\[
\frac{c^\top\Psi(y)}{\rho\bigl(\Psi(y)\bigr)} =\frac{1/(\one^\top y)}{\rho(y)/(\one^\top y)} =\frac{1}{\rho(y)} .
\]
Maximizing $c^\top x/\rho(x)$ over $\Delta_C^+$ is therefore equivalent to minimizing $\rho$ over $F$, with optimal solutions related by $\Phi$ and $\Psi$; in particular $x^\star=\Psi(y^\star)$ is optimal for~\eqref{prob:ratio-generic} whenever $y^\star$ is optimal for~\eqref{prob:ratio-transf}.

\emph{Convexity.} The objective of~\eqref{prob:ratio-transf} is convex by hypothesis, and its feasible set is the intersection of the convex cone $C$ with a hyperplane.
\end{proof}

\section*{Statements and Declarations}

\paragraph{Funding} The authors did not receive support from any organization for the submitted work.

\paragraph{Competing interests} The authors have no relevant financial or non-financial interests to disclose.

\paragraph{Data availability} The datasets and results are available at \href{https://www.francescocesarone.com/data-sets}{francescocesarone.com} \citep{Cesarone2026data}; the raw prices are from FactSet, subject to its license terms.

\paragraph{Code availability} The MATLAB code is available from the corresponding author on reasonable request.

\bibliographystyle{elsarticle-harv}
\bibliography{Bib_CCGL_semivol}

\clearpage
\setcounter{section}{0}\setcounter{subsection}{0}
\setcounter{table}{0}\setcounter{figure}{0}\setcounter{equation}{0}
\counterwithout{equation}{section}
\counterwithout{table}{section}
\counterwithout{figure}{section}
\renewcommand{\thesection}{S.\arabic{section}}
\renewcommand{\thesubsection}{S.\arabic{section}.\arabic{subsection}}
\renewcommand{\thetable}{S.\arabic{table}}
\renewcommand{\thefigure}{S.\arabic{figure}}
\renewcommand{\theequation}{S.\arabic{equation}}
\renewcommand{\theHsection}{SM.\thesection}
\renewcommand{\theHsubsection}{SM.\thesubsection}
\renewcommand{\theHtable}{SM.\thetable}
\renewcommand{\theHfigure}{SM.\thefigure}
\renewcommand{\theHequation}{SM.\theequation}
\newcommand{\SVol}{\ensuremath{\varsigma}}
\newcommand{\LPM}{\ensuremath{\mathrm{LPM}}}
\newcommand{\sig}{\ensuremath{^{*}}}
\newcommand{\RQ}{\paragraph*{Research question.}}
\newcommand{\ME}{\paragraph*{Methodology.}}
\newcommand{\RE}{\paragraph*{Results.}}
\renewcommand{\bottomfraction}{0.8}
\renewcommand{\textfraction}{0.07}
\emergencystretch=3em
\setlength{\parskip}{0.35em}
\setlength{\tabcolsep}{5pt}
\renewcommand{\arraystretch}{1.08}
\phantomsection
\pdfbookmark[0]{Supplementary Material}{supplementary}
\begin{center}
{\LARGE\bfseries Supplementary Material}\\[0.6em]
{\large Semivariance and Semivolatility in Portfolio Selection: A Unified Framework for Downside-Risk Portfolio Models}
\end{center}
\vspace{1em}
\begin{quote}\small
\noindent\textbf{Abstract.}
\noindent This supplement reports four additional analyses of the out-of-sample experiment described in the main text.
They address three questions that the performance tables of the paper leave open: whether the reported differences are statistically distinguishable from sampling noise, how far apart the portfolios of models with indistinguishable performance are, and whether the differences survive alternative design choices and rebalancing calendars.
Part~I (Sections~\ref{S-sec:signif} and~\ref{S-sec:anatomy}) works on the out-of-sample series of the main experiment; Part~II (Sections~\ref{S-sec:design} and~\ref{S-sec:timing}) re-estimates the models under alternative designs and shifted rebalancing calendars.
Every analysis covers both investment universes of the main text, and Section~\ref{S-sec:summary} collects the results.
Section~\ref{S-sec:theory} maps the theoretical results of the main text to their antecedents and reports the numerical accuracy of the optimal portfolios.
Estimators are collected in Appendix~\ref{S-app:est}.
\end{quote}

\etocdepthtag.toc{mtsm}
\etocsettagdepth{mtmain}{none}
\etocsettagdepth{mtsm}{subsection}
\tableofcontents

\section{Scope and common experimental setting}
\label{S-sec:scope}

\subsection{Design}

All analyses in this supplement start from the rolling-window experiment of the main text and, unless a section states otherwise, use its configuration: learning window $D_L=504$ trading days, holding window $D_H=21$ trading days, no cap on individual weights, proportional transaction costs of $50$ basis points charged on the rebalancing turnover, and a static investable universe.
As in the main text, the universes consist of current index constituents, whose survivorship and look-ahead bias favors every strategy relative to the index, so all index-relative figures in this supplement are descriptive; both the asset prices and the market indexes are total-return series from FactSet, so that dividends are included throughout.
The cost level of $50$ basis points follows \citet{S-demiguel2009optimal}; the turnover of every model is reported in Tables~4 and~5 of the main text.
The optimal weights do not depend on the cost level, so the same backtest gives the net returns at any level, and Table~\ref{S-tab:costs} reports the Sharpe ratio of every model at $0$, $25$, $50$ and $100$ basis points.
Moving from $0$ to $100$ basis points lowers the Sharpe ratio of the equally weighted portfolio by about $0.03$ (DowJones) and $0.04$ (Nasdaq100) and that of mean--variance by $0.12$ to $0.15$, so the higher-turnover models lose ground to the equally weighted portfolio as costs rise, whereas the difference between RP-SVol$_{\mu}$ and RP-Vol changes by at most $0.001$.
The seventeen portfolio models are those of Table~3 of the main text, and the market index of each universe is the total-return version of the index whose constituents form it (Table~2 of the main text); the notation for thresholds follows the paper, with subscripts $\mu$, $0$ and $R^I$ denoting the below-mean, below-zero and below-benchmark variants.

The two universes, DowJones and Nasdaq100, span March 2008 to November 2026 (DowJones) and October 2012 to November 2026 (Nasdaq100) and are constructed with the complete-series filter of the main text (Table~\ref{S-tab:universes}).
As in the main text, the risk-free rate is zero; a positive rate $r_f$ would lower each Sharpe ratio by $r_f/\sigma^{\mathrm{ann}}$, which changes the difference between two models little only when their volatilities are similar.

\begin{table}[htbp]
\centering
\caption{Investment universes. $n$ is the number of assets with a complete price
series over the sample period, $N^{\mathrm{out}}$ the number of out-of-sample
daily observations and $Q$ the number of rolling windows.}
\label{S-tab:universes}
\begin{adjustbox}{max width=\textwidth,max totalheight=0.8\textheight}
\begin{tabular}{lrrr}
\toprule
Universe & $n$ & $N^{\mathrm{out}}$ & $Q$ \\
\midrule
DowJones  & 30 & 4179 & 199 \\
Nasdaq100 & 84 & 3024 & 144 \\
\bottomrule
\end{tabular}
\end{adjustbox}
\end{table}

\begin{table}[htbp]
\centering
\caption{Annualized out-of-sample Sharpe ratio of the seventeen models at proportional transaction costs $c_{\mathrm{tc}}$ of $0$, $25$, $50$ and $100$ basis points (natural units; $50$ basis points is the level of the main text). The best model at each cost level is in bold, ties at three decimals being broken on unrounded values.}
\label{S-tab:costs}
\begin{adjustbox}{max width=\textwidth,max totalheight=0.8\textheight}
\begin{tabular}{lrrrrrrrr}
\toprule
 & \multicolumn{4}{c}{DowJones} & \multicolumn{4}{c}{Nasdaq100} \\
\cmidrule(lr){2-5}\cmidrule(lr){6-9}
$c_{\mathrm{tc}}$ (bp) & 0 & 25 & 50 & 100 & 0 & 25 & 50 & 100 \\
\midrule
EW & 1.065 & 1.057 & 1.049 & \textbf{1.034} & 1.174 & \textbf{1.165} & \textbf{1.156} & \textbf{1.138} \\
MV & 0.833 & 0.804 & 0.774 & 0.715 & 0.941 & 0.905 & 0.869 & 0.796 \\
MSV$_{0}$ & 0.860 & 0.824 & 0.787 & 0.715 & 0.970 & 0.925 & 0.880 & 0.789 \\
MSV$_{\mu}$ & 0.832 & 0.800 & 0.767 & 0.701 & 0.930 & 0.889 & 0.848 & 0.766 \\
MSV$_{R^{I}}$ & 0.893 & 0.883 & 0.872 & 0.851 & 1.049 & 1.032 & 1.016 & 0.983 \\
MaxSharpe & 1.008 & 0.947 & 0.885 & 0.762 & 0.952 & 0.888 & 0.824 & 0.696 \\
MaxSortino$_{\mu}$ & 0.981 & 0.921 & 0.860 & 0.739 & 0.932 & 0.872 & 0.811 & 0.689 \\
MaxSortino$_{0}$ & 0.976 & 0.916 & 0.855 & 0.734 & 0.924 & 0.863 & 0.803 & 0.681 \\
MaxSortino$_{R^{I}}$ & 0.944 & 0.920 & 0.896 & 0.847 & 1.066 & 1.044 & 1.021 & 0.975 \\
RP-Vol & 1.058 & 1.049 & 1.040 & 1.023 & 1.148 & 1.137 & 1.127 & 1.107 \\
RP-SVol$_{\mu}$ & 1.057 & 1.048 & 1.039 & 1.021 & 1.145 & 1.134 & 1.124 & 1.103 \\
RP-SVol$_{0}$ & 1.058 & 1.049 & 1.041 & 1.024 & 1.145 & 1.135 & 1.125 & 1.105 \\
RP-SVol$_{R^{I}}$ & 0.909 & 0.900 & 0.891 & 0.873 & 1.072 & 1.062 & 1.052 & 1.033 \\
MDP-Vol & 1.085 & 1.057 & 1.028 & 0.971 & \textbf{1.176} & 1.142 & 1.109 & 1.041 \\
MDP-SVol$_{\mu}$ & 1.142 & 1.107 & 1.071 & 0.999 & 1.076 & 1.036 & 0.997 & 0.917 \\
MDP-SVol$_{0}$ & \textbf{1.160} & \textbf{1.125} & \textbf{1.089} & 1.018 & 1.081 & 1.043 & 1.004 & 0.927 \\
MDP-SVol$_{R^{I}}$ & 0.893 & 0.883 & 0.872 & 0.851 & 1.052 & 1.035 & 1.019 & 0.987 \\
\bottomrule
\end{tabular}
\end{adjustbox}
\end{table}

\begin{figure}[htbp]
\centering
\includegraphics[width=\textwidth]{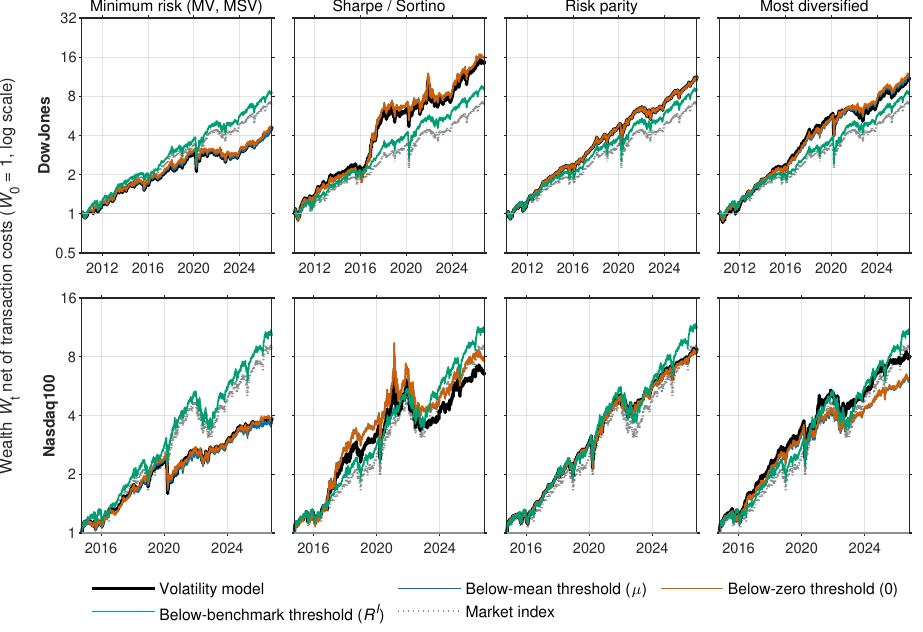}
\caption{Out-of-sample wealth $W_t$ net of transaction costs ($W_0=1$, logarithmic scale) on DowJones and Nasdaq100. Each column is a class of models: minimum risk (MV and MSV), Sharpe and Sortino ratio, risk parity, most-diversified portfolio. In each panel the black line is the volatility-based model and the colored lines are its below-mean ($\mu$), below-zero ($0$) and below-benchmark ($R^I$) variants; the dotted line is the market index. The vertical scale is common within each row. Where curves overlap, as in risk parity, the black line shows only at the edges.}
\label{S-fig:wealth}
\end{figure}

Figure~\ref{S-fig:wealth} shows the out-of-sample wealth of the volatility model of each class, of its three downside variants and of the market index, net of transaction costs, in the two universes.
Within each class the volatility model and its below-mean and below-zero variants stay close over the whole period, except in the most-diversified class on Nasdaq100, where MDP-Vol ends about $30\%$ above its two variants ($8.01$ against $6.13$ and $6.20$).
In risk parity their curves can hardly be told apart, in line with the distances of Table~\ref{S-tab:anatomy}.
The below-benchmark variant follows a path of its own, closer in shape to that of the market index, and in the minimum-risk class it ends well above the other three variants on both universes.
On Nasdaq100, for instance, the final wealth ranges from $3.70$ for MSV$_{\mu}$ to $11.43$ for RP-SVol$_{R^I}$, against $8.82$ for the market index; for the reasons given above, this comparison is descriptive.

The experimental protocol of the main text follows \citet{S-demiguel2009optimal}, and two of the analyses below are taken directly from that paper: the test on Sharpe ratio differences against the equally weighted portfolio (Section~\ref{S-sec:signif}) and the variation of the estimation window (Section~\ref{S-sec:design}).

\subsection{Inference}
\label{S-sec:inference}

Two conventions recur throughout.
Since the out-of-sample series are concatenated holding periods, all variances are estimated in heteroskedasticity- and autocorrelation-consistent (HAC) form \citep{S-newey1987simple,S-andrews1991heteroskedasticity} with a Bartlett kernel, whose bandwidth (Appendix~\ref{S-app:est}) equals $9$ (DowJones) and $8$ (Nasdaq100) for series of about $4179$ and $3024$ observations.
Wherever a $p$-value is required it is obtained by circular block bootstrap \citep{S-politis1992circular} with blocks of $21$ trading days and $B=1999$ replications, using the same resampling indices for every pairwise comparison within an analysis and fixed random seeds, so that reported $p$-values are exactly reproducible.
All tests in this supplement concern the Sharpe ratio, as do those of Table~6 of the main text.

\part*{Part I: Analyses of the out-of-sample series}
\addcontentsline{toc}{section}{Part I: Analyses of the out-of-sample series}

\section{Statistical significance of performance differences}
\label{S-sec:signif}

\RQ We ask whether the differences in Sharpe ratios between the portfolio models and the natural benchmarks (the equally weighted portfolio, the minimum-variance portfolio and the market index) are distinguishable from sampling noise.

\ME For each universe we stack the moment conditions $Y=[\,X,\;X^2\,]$, where $X$ is the $N^{\mathrm{out}}\times 18$ matrix of out-of-sample returns of the $18$ series, and estimate a single HAC covariance matrix $\widehat V$ of $Y$.
For a given pair of series the difference of two Sharpe ratios and its standard error follow from the delta method as a $4\times4$ quadratic form in $\widehat V$; the gradients are given in Appendix~\ref{S-app:delta}.
This is the test of \citet{S-jobson1981performance} in the correction of \citet{S-memmel2003performance}, in the heteroskedasticity- and autocorrelation-robust version of \citet{S-ledoit2008robust}.
The reported $p$-values are studentized block-bootstrap $p$-values: at each replication the means and the HAC matrix are recomputed on the bootstrap sample and the $p$-value is $(1+c)/(B+1)$, where $c$ is the number of replications with $|(\hat d^{*}-\hat d)/\widehat{se}^{*}|\ge|\hat d/\widehat{se}|$.
The confidence intervals are equal-tailed percentile-$t$ intervals from the same replications, whereas the $p$-values are symmetric, so that an interval can exclude zero while the unadjusted $p$-value slightly exceeds $0.05$.
Because seventeen comparisons are performed against each benchmark, the significance marks of Tables~\ref{S-tab:signif} and~\ref{S-tab:signifcounts} are based on unadjusted $p$-values and must be read with that multiplicity in mind.
The within-family tests of Table~6 of the main text report $p$-values adjusted by the procedure of \citet{S-holm1979simple}, which controls the family-wise error rate.
Two related procedures are \emph{not} implemented here: the deflated Sharpe ratio of \citet{S-bailey2014deflated} and the tests of superior predictive ability of \citet{S-white2000reality} and \citet{S-hansen2005test}, which would address the complementary question of whether the best of seventeen models beats the benchmark once the selection is taken into account.

\RE Table~\ref{S-tab:signif} reports the differences in annualized Sharpe ratio relative to the minimum-variance portfolio, and Table~\ref{S-tab:signifcounts} summarizes the counts of significant comparisons against the three benchmarks.

\begin{table}[htbp]
\centering
\caption{Difference in annualized Sharpe ratio relative to MV.
\sig\ denotes unadjusted $p<0.05$ under the studentized block bootstrap
($B=1999$). Natural units.}
\label{S-tab:signif}
\small
\begin{adjustbox}{max width=\textwidth,max totalheight=0.8\textheight}
\begin{tabular}{lrr}
\toprule
 & DowJones & Nasdaq100 \\
\midrule
EW                     & $+0.27$ & $+0.29$ \\
MSV$_{0}$              & $+0.01$ & $+0.01$ \\
MSV$_{\mu}$            & $-0.01$ & $-0.02$ \\
MSV$_{R^{I}}$          & $+0.10$ & $+0.15$ \\
MaxSharpe              & $+0.11$ & $-0.04$ \\
MaxSortino$_{\mu}$     & $+0.09$ & $-0.06$ \\
MaxSortino$_{0}$       & $+0.08$ & $-0.07$ \\
MaxSortino$_{R^{I}}$   & $+0.12$ & $+0.15$ \\
RP-Vol                 & $+0.27$\sig & $+0.26$ \\
RP-SVol$_{\mu}$        & $+0.26$\sig & $+0.25$ \\
RP-SVol$_{0}$          & $+0.27$\sig & $+0.26$ \\
RP-SVol$_{R^{I}}$      & $+0.12$ & $+0.18$ \\
MDP-Vol                & $+0.25$\sig & $+0.24$ \\
MDP-SVol$_{\mu}$       & $+0.30$\sig & $+0.13$ \\
MDP-SVol$_{0}$         & $+0.31$\sig & $+0.14$ \\
MDP-SVol$_{R^{I}}$     & $+0.10$ & $+0.15$ \\
Market index           & $+0.03$ & $+0.07$ \\
\bottomrule
\end{tabular}
\end{adjustbox}
\end{table}

\begin{table}[htbp]
\centering
\caption{Number of series (out of seventeen) with a significantly higher or
lower Sharpe ratio than the reference series of each column (EW, MV, market index) at the $5\%$ level, unadjusted bootstrap
$p$-values.}
\label{S-tab:signifcounts}
\begin{adjustbox}{max width=\textwidth,max totalheight=0.8\textheight}
\begin{tabular}{lcccccc}
\toprule
 & \multicolumn{2}{c}{vs EW} & \multicolumn{2}{c}{vs MV} & \multicolumn{2}{c}{vs index} \\
\cmidrule(lr){2-3}\cmidrule(lr){4-5}\cmidrule(lr){6-7}
Universe & higher & lower & higher & lower & higher & lower \\
\midrule
DowJones  & 0 & 5 & 6 & 0 & 11 & 0 \\
Nasdaq100 & 0 & 1 & 0 & 0 & 5  & 0 \\
\bottomrule
\end{tabular}
\end{adjustbox}
\end{table}

In this section, unless stated otherwise, ``significant'' refers to unadjusted bootstrap $p$-values at the $5\%$ level.

First, the differences between mean--variance and the two minimum-semivariance models MSV$_0$ (below zero) and MSV$_\mu$ (below mean) are not statistically distinguishable from zero.
The two rows of Table~\ref{S-tab:signif} range from $-0.02$ to $+0.01$ and are never significant, and neither is the row of MSV$_{R^I}$, whose benchmark threshold changes the objective.

Second, on DowJones mean--variance is outperformed significantly only by risk budgeting and by the diversification-ratio models.
There the three risk-parity variants with the volatility, below-mean and below-zero risk measures, MDP-Vol, MDP-SVol$_\mu$ and MDP-SVol$_0$ are significant, while the equally weighted portfolio ($+0.27$, $p=0.066$) narrowly misses the $5\%$ level; on Nasdaq100 none is significant (the largest difference among these models, $+0.26$ for RP-Vol, has $p=0.115$).
Within risk budgeting the volatility and semivolatility variants have almost identical magnitudes ($+0.266$ for RP-Vol and $+0.265$ for RP-SVol$_\mu$ on DowJones), whereas those of the diversification-ratio models differ by up to about $0.11$ ($+0.24$ for MDP-Vol against $+0.13$ for MDP-SVol$_\mu$ on Nasdaq100).
The direct comparison of each downside model with its variance-based counterpart, and of the three thresholds within each family, is reported in Table~6 of the main text, where no difference between a downside model and its variance-based counterpart is significant and the confidence intervals are narrow only for risk budgeting.

The equivalence of RP-SVol$_{\mu}$ and RP-Vol in the main text rests on a block length of $21$ days and a HAC bandwidth of $9$ ($8$ on Nasdaq100).
Table~\ref{S-tab:sensblock} repeats the computation with blocks of $63$ and $126$ days and with the bandwidth halved and doubled: the smallest equivalence margin varies by at most $0.0004$ across the five settings in each universe and stays below the cost margin $\delta$ in all ten cases.
In the other three classes the same changes leave the smallest margins between $0.045$ and $0.219$, so they remain far above $\delta$.
Since $\delta$ is proportional to $c_{\mathrm{tc}}$, at $25$ basis points it falls to about $0.009$ and $0.010$, still above $\delta^{*}$ on DowJones in every setting but below it on Nasdaq100 in all five, so that there the equivalence would not be declared at that cost level.
Under the same five settings the risk-parity difference between the below-mean and benchmark thresholds (Panel~C of Table~6 of the main text) has per-panel Holm $p$-values between $0.002$ and $0.006$ on DowJones, and equal to one on Nasdaq100.

\begin{table}[htbp]
\centering
\caption{Smallest margin $\delta^{*}$ at which the two one-sided tests procedure ($90\%$ studentized circular block-bootstrap interval, $B=1999$) declares RP-SVol$_{\mu}$ and RP-Vol equivalent in Sharpe ratio, for different block lengths and HAC bandwidths $L$ (on Nasdaq100 the baseline bandwidth is $8$, halved to $4$ and doubled to $16$). The last row gives the cost-based margin $\delta=12\,c_{\mathrm{tc}}\,\mathrm{Turn}/\sigma^{\mathrm{ann}}$ of the main text, with the turnover $\mathrm{Turn}$, as a fraction, and the annualized volatility $\sigma^{\mathrm{ann}}$ of RP-Vol.}
\label{S-tab:sensblock}
\small
\begin{adjustbox}{max width=\textwidth,max totalheight=0.8\textheight}
\begin{tabular}{lcc}
\toprule
Block, $L$ & DowJones & Nasdaq100 \\
\midrule
$21$, $9$ (baseline) & 0.0066 & 0.0110 \\
$63$, $9$            & 0.0068 & 0.0110 \\
$126$, $9$           & 0.0065 & 0.0107 \\
$21$, $4$            & 0.0065 & 0.0110 \\
$21$, $18$           & 0.0065 & 0.0109 \\
\midrule
$\delta$             & 0.0175 & 0.0205 \\
\bottomrule
\end{tabular}
\end{adjustbox}
\end{table}

Third, the equally weighted portfolio is not significantly outperformed in either universe, while it significantly outperforms the market index on both universes and the four benchmark-threshold models on DowJones.
This is the finding of \citet{S-demiguel2009optimal}; for the opposite view, based on sample length and estimation quality, see \citet{S-kritzman2010defense}.

\section{Portfolio anatomy: how distinct are the models?}
\label{S-sec:anatomy}

\RQ The performance of the mean--semivariance and mean--variance models cannot be distinguished at this sample length.
We ask whether this is because the two portfolios are nearly identical, or because they are different and the data cannot separate them.

\ME Using the optimal weights of every rolling window, we compute for each pair of models the average $L_1$ distance between the weight vectors.
The $L_1$ distance is the same quantity that defines turnover in the main text and can be read as the trading that would be required to move from one model to the other, counting both purchases and sales (two-way convention), so that a distance $d$ corresponds to a fraction $d/2$ of the portfolio value moved one way; it ranges from $0$ for identical portfolios to $2$ for disjoint ones.
The exercise belongs to the literature on estimation error in optimal portfolios \citep{S-michaud1989markowitz,S-best1991sensitivity,S-chopra1993effect}, which documents how strongly mean--variance weights respond to small perturbations of the inputs, and to \citet{S-jagannathan2003risk}, who show that weight constraints act as implicit shrinkage.
The distance measure itself is our own construction.

\RE Table~\ref{S-tab:anatomy} reports the average $L_1$ distance for the pairs of interest.

\begin{table}[htbp]
\centering
\caption{Average $L_1$ distance between optimal weight vectors, averaged across
rolling windows ($0$ = identical portfolios, $2$ = maximal distance).}
\label{S-tab:anatomy}
\small
\setlength{\tabcolsep}{4pt}
\begin{adjustbox}{max width=\textwidth,max totalheight=0.8\textheight}
\begin{tabular}{lrr}
\toprule
Pair & DowJones & Nasdaq100 \\
\midrule
RP-Vol / RP-SVol$_{\mu}$           & 0.030 & 0.042 \\
MSV$_{0}$ / MSV$_{\mu}$            & 0.190 & 0.267 \\
MaxSharpe / MaxSortino$_{\mu}$     & 0.223 & 0.342 \\
MV / MSV$_{\mu}$                   & 0.252 & 0.370 \\
MV / MSV$_{0}$                     & 0.336 & 0.473 \\
MDP-Vol / MDP-SVol$_{\mu}$         & 0.347 & 0.465 \\
\bottomrule
\end{tabular}
\end{adjustbox}
\end{table}

Within risk budgeting, volatility and semivolatility deliver very close portfolios on our data: an $L_1$ distance of $0.03$ to $0.04$ means trades of $3$ to $4\%$ of the portfolio value counting purchases and sales, that is, at most about $2\%$ one way.
The nearly identical performance of RP-Vol and RP-SVol$_\mu$ therefore reflects the empirical proximity of the two portfolios rather than two different allocations with similar outcomes.
This proximity is a feature of the observed returns, not an algebraic identity: the two portfolios coincide when the return of every long-only portfolio is symmetric about its mean (Section~2.2 of the main text), and can differ under asymmetric returns.
For MV against MSV$_0$, and for MDP-Vol against MDP-SVol$_\mu$, the distance is an order of magnitude larger, between $0.34$ and $0.47$, and yet the performance of these pairs cannot be distinguished at this sample length; whether a distance of this size exceeds the sampling variability of the estimated weights is not assessed in this supplement.

\part*{Part II: Analyses that re-estimate the models}
\addcontentsline{toc}{section}{Part II: Analyses that re-estimate the models}

\noindent The two analyses below solve the optimization problems again, under alternative designs and on shifted rebalancing calendars.

\section{Sensitivity to design parameters}
\label{S-sec:design}

\RQ We ask whether the ordering of the models depends on the triple $(D_L,D_H,\mathrm{ub})$ adopted in the main text, where $\mathrm{ub}$ is the upper bound on the individual weights of the minimum-risk, reward-to-risk and diversification-ratio models ($x_i\le\mathrm{ub}$; $\mathrm{ub}=1$ in the main text).

\ME We run a one-factor-at-a-time sensitivity analysis around the baseline configuration, with $D_L\in\{252,504,756\}$, $D_H\in\{21,63,126\}$ and $\mathrm{ub}\in\{0.05,0.10,0.20,1\}$, which gives eight configurations per universe; the full factorial design over the same grids ($36$ configurations) is implemented but not run.
For each configuration the entire experiment is re-run for all seventeen models, and we report the performance measures together with the Spearman rank correlation between the resulting ordering of the seventeen models and the baseline ordering.
Configurations with $\mathrm{ub}<1/n$ would be infeasible on a static universe; with $n\ge30$ none of those considered is.
Varying the estimation window as a robustness check follows \citet{S-demiguel2009optimal}; treating the weight cap as a design parameter rather than a technical constraint follows \citet{S-jagannathan2003risk}.
The rebalancing frequency is added here as a third axis.
The caps are imposed on the optimized models only: they leave the equally weighted portfolio unchanged, since $1/n\le1/30$ never reaches $0.05$, and they are not imposed on the four risk-budgeting portfolios, whose programs have no upper bound on the weights and which are therefore identical in the four configurations that differ only in $\mathrm{ub}$.
In the alternative configurations the solver never fell back to the equally weighted allocation, on either universe.
A linear program shows that $\Delta_C^+$ is not empty in any window, including those with $\mathrm{ub}=0.05$, so that Lemma~2 of the main text applies throughout.
In the calendar analysis of Section~\ref{S-sec:timing} no model fell back to the equally weighted allocation in any window, under any of the seven offsets, on either universe.

\RE Table~\ref{S-tab:design} reports the rank correlation between each alternative configuration and the baseline.

\begin{table}[htbp]
\centering
\caption{Spearman rank correlation between the Sharpe ratio ordering of the
seventeen models under each configuration and under the baseline
$(D_L,D_H,\mathrm{ub})=(504,21,1)$. Each configuration changes one parameter only.
The last three rows count the configurations, baseline included, in which the
stated Sharpe-ratio inequality holds.}
\label{S-tab:design}
\small
\begin{adjustbox}{max width=\textwidth,max totalheight=0.8\textheight}
\begin{tabular}{lrr}
\toprule
Configuration & DowJones & Nasdaq100 \\
\midrule
$D_L=252$  & 0.85 & 0.95 \\
$D_L=756$  & 0.95 & 0.99 \\
$D_H=63$   & 0.95 & 0.95 \\
$D_H=126$  & 0.92 & 0.87 \\
$\mathrm{ub}=0.05$  & 0.69 & 0.51 \\
$\mathrm{ub}=0.10$  & 0.73 & 0.88 \\
$\mathrm{ub}=0.20$  & 0.91 & 0.95 \\
\midrule
\multicolumn{3}{l}{\textit{Within-class sign counts (of the eight configurations)}} \\
RP-SVol$_\mu$ $<$ RP-Vol & 8 & 8 \\
MDP-SVol$_\mu$ $<$ MDP-Vol & 0 & 8 \\
MSV$_0$ $>$ MV & 6 & 5 \\
\bottomrule
\end{tabular}
\end{adjustbox}
\end{table}

The ordering of the models is robust to the learning window and to the rebalancing frequency on both universes, and less so to the weight cap.
On Nasdaq100 the correlation stays above $0.84$ in every configuration except the tightest cap ($0.51$).
On DowJones, the smallest universe, the correlation lies between $0.69$ and $0.95$ in all seven alternatives, because most models there are separated by a few hundredths of Sharpe ratio.
With semiannual rebalancing, which leaves only $33$ holding periods on DowJones and $24$ on Nasdaq100, the correlations are $0.92$ and $0.87$.
A different learning window also shifts the out-of-sample period, which starts $D_L$ days after the beginning of the sample.
Computed on the dates common to the three windows (from March 2011 to October 2026, $3927$ days, on DowJones and from October 2015 to October 2026, $2772$ days, on Nasdaq100), the correlations with the baseline are $0.88$ and $0.90$ for $D_L=252$ and $0.98$ and $0.99$ for $D_L=756$, on DowJones and Nasdaq100 respectively.
On the common period the correlations change little, the largest move being that for $D_L=252$ on Nasdaq100 (from $0.95$ to $0.90$), so that the different sample does not account for the lower values in Table~\ref{S-tab:design}.
Among the design parameters, the cap at $5\%$ is the most disruptive on both universes; it binds on the concentrated models and moves them toward the diversified ones.

The comparison within each class, on which the paper rests, is more stable than the full ordering.
Over the $16$ configurations (eight in each of two universes), RP-SVol$_\mu$ has a lower Sharpe ratio than RP-Vol in all $16$, and MDP-SVol$_\mu$ has a lower Sharpe ratio than MDP-Vol in $8$, all on Nasdaq100.
These counts describe how often a sign is preserved and are not independent replications: the configurations share the universe, the sample period and most of the estimation windows, so their outcomes are strongly dependent.
The comparison between MSV$_0$ and MV is positive in $11$ configurations against $5$, and its sign depends on the configuration: it is positive in six of the eight configurations on DowJones and in five on Nasdaq100.
The risk-budgeting and diversification-ratio comparisons of Section~\ref{S-sec:signif} therefore keep their sign in all the alternative triples on both universes, the diversification-ratio one with opposite signs on DowJones and Nasdaq100, whereas that of MSV$_0$ against MV does not; their significance is not re-tested here.

\section{Rebalancing-date luck}
\label{S-sec:timing}

\RQ The rolling scheme starts on the first day of the sample and rebalances every $D_H=21$ days, but there are $21$ equally legitimate starting points.
We ask how much of the measured difference between models survives a change of the rebalancing calendar.

\ME The experiment is repeated for a set of offsets $o$, each time on the sample $\{o+1,\dots,T\}$, and we record the dispersion of the performance measures across offsets together with the stability of the ordering, measured by the Spearman correlation against the baseline offset and by the identity of the best-performing model.
Shifting the start also changes the number of windows slightly, so levels are not exactly comparable across offsets and the comparison is made on orderings.
Since the offsets move the start by at most $20$ trading days out of about $4700$ on DowJones and $3500$ on Nasdaq100, the seven series cover nearly the same period, and the restriction to common dates used for $D_L$ in Section~\ref{S-sec:design} is not needed here.
The phenomenon is documented for smart-beta indexes by \citet{S-hoffstein2020rebalance}, who show that the choice of rebalancing date alone generates performance dispersion of the same order as the differences between strategies; applying this rebalancing-offset analysis to a comparison of optimization models is the variant proposed here.
We use seven offsets, $o\in\{0,3,7,10,13,17,20\}$, which span the monthly cycle.

\RE Table~\ref{S-tab:timing} summarizes the dispersion across the seven calendars.

\begin{table}[htbp]
\centering
\caption{Rebalancing-date luck across seven offsets. $r^{\mathrm{S}}_{\min}$ is the
lowest Spearman correlation between the Sharpe ratio ordering under an offset
and under the baseline; ``best models'' is the number of distinct models ranked
first; ``range'' is the maximum minus the minimum Sharpe ratio across offsets.
The last three rows count the offsets, out of seven, in which the first model
has the higher Sharpe ratio.}
\label{S-tab:timing}
\small
\setlength{\tabcolsep}{4pt}
\begin{adjustbox}{max width=\textwidth,max totalheight=0.8\textheight}
\begin{tabular}{lrr}
\toprule
 & DowJones & Nasdaq100 \\
\midrule
$r^{\mathrm{S}}_{\min}$                  & 0.94 & 0.95 \\
Best models                    & 1 & 1 \\
Largest range                  & 0.082 & 0.115 \\
Range of EW                    & 0.012 & 0.044 \\
\midrule
MSV$_0$ vs MV                  & 6 & 7 \\
RP-SVol$_\mu$ vs RP-Vol        & 1 & 0 \\
MDP-SVol$_\mu$ vs MDP-Vol      & 7 & 0 \\
\bottomrule
\end{tabular}
\end{adjustbox}
\end{table}

The rebalancing calendar moves the level of every model but leaves the comparisons within each class almost unchanged.
On both universes the ordering is essentially fixed, with $r^{\mathrm{S}}_{\min}$ equal to $0.94$ on DowJones and $0.95$ on Nasdaq100, and the same model ranks first under every calendar: MDP-SVol$_0$ on DowJones and the equally weighted portfolio on Nasdaq100.
The Sharpe ratio of a single model varies across calendars by up to $0.08$ on DowJones and $0.11$ on Nasdaq100.
The largest ranges belong to the concentrated models, namely MSV$_\mu$ and MSV$_0$ on DowJones and MaxSortino$_\mu$ and MaxSortino$_0$ on Nasdaq100, while the equally weighted and risk-budgeting portfolios vary by at most $0.05$.
These ranges are comparable to, or larger than, the within-class differences that Section~\ref{S-sec:signif} examines.

The comparisons within each class survive nonetheless, because the two members of a pair are exposed to the same calendar.
RP-SVol$_\mu$ is below RP-Vol under $13$ of the $14$ calendar-universe combinations, and MDP-SVol$_\mu$ below MDP-Vol in $7$, all on Nasdaq100, counts that, as in Section~\ref{S-sec:design}, are not independent replications.
The sign of the MSV$_0$ against MV difference is positive under all seven calendars on Nasdaq100 and under six of the seven on DowJones.
A single calendar is therefore adequate for the within-class comparisons on which the paper rests; the identity of the model ranked first does not change with the calendar on either universe, although the Sharpe ratio of a single model moves by up to $0.11$.

\section{Summary of results}
\label{S-sec:summary}

The four analyses support the following statements.

\begin{enumerate}[leftmargin=1.6em,itemsep=0.35em]
\item \textbf{Within each model class, the downside and the variance-based
      formulations cannot be distinguished out of sample at this sample length after the multiplicity adjustment}
      (Section~\ref{S-sec:signif}). In risk budgeting the two select very close
      portfolios on our data (Section~\ref{S-sec:anatomy}); for MV against MSV$_0$ and
      for MDP-Vol against MDP-SVol$_\mu$ the average $L_1$ distance between
      the two portfolios is between $0.34$ and $0.47$.
\item \textbf{The within-class comparisons are robust to the design; the full
      ordering is not} (Sections~\ref{S-sec:design} and~\ref{S-sec:timing}). Across
      $16$ design configurations and $14$ rebalancing calendars the
      semivolatility risk-budgeting models stay slightly below their
      volatility counterparts in almost every case, the diversification-ratio
      ones are below on Nasdaq100 and above on DowJones, the
      sign of MSV$_0$ against MV is not stable across design choices, and the full
      ordering of the models moves with the weight cap.
\item \textbf{The equally weighted portfolio is not significantly outperformed in
      either universe}
      (Section~\ref{S-sec:signif}).
\end{enumerate}

\section{Theoretical results and numerical accuracy}
\label{S-sec:theory}

\subsection{Antecedents of the theoretical results}
\label{S-sec:antecedents}

Table~\ref{S-tab:antecedents} lists the theoretical results of Sections~2--3 and of Appendices~A--B of the main text, with the hypotheses under which each is stated, its closest antecedent in the literature and what the paper adds to it.
A result is classified as \emph{recalled} when it is known in the form used, as a \emph{specialization} when it is a known general result instantiated for below-mean semivolatility or for the threshold functional $\rho_\tau$, as an \emph{extension} when it adds new statements to a known result, and as \emph{new} when we are not aware of a previous statement.
Most of the analytical tools are recalled, specialized or extended; the new elements are the characterization of the additivity of below-mean semivolatility, the observation that non-degeneracy relative to a random threshold does not follow from its below-mean form, and the use of the degree and the threshold as the two axes that organize the models.

\begin{table}[htbp]
\centering
\caption{Theoretical results of the main text: hypotheses, closest antecedent,
status and contribution of the paper. Numbering of results, equations and
assumptions as in the main text; Assumption~1 is the below-mean
non-degeneracy condition, Assumption~2 its version relative to a target $\tau$.}
\label{S-tab:antecedents}
\footnotesize
\setlength{\tabcolsep}{3pt}
\renewcommand{\arraystretch}{1.05}
\begin{adjustbox}{max width=\textwidth,max totalheight=0.8\textheight}
\begin{tabular}{>{\raggedright\arraybackslash}p{3.3cm}
                >{\raggedright\arraybackslash}p{2.45cm}
                >{\raggedright\arraybackslash}p{3.05cm}
                >{\raggedright\arraybackslash}p{2.3cm}
                >{\raggedright\arraybackslash}p{4.2cm}}
\toprule
Result & Hypotheses & Antecedent & Status & What the paper adds \\
\midrule
Prop.~1: properties of below-mean semivolatility
& $X\in L^2$
& \citet{S-ogryczak1999from,S-ogryczak2001consistency}; \citet{S-fischer2003risk}
& Recalled
& Short proof in our notation. \\
Prop.~2: equivalent forms of Assumption~1
& $R\in(L^2)^n$
& \citet{S-ararat2021mad}, form~(ii)
& Extension
& Adds forms~(iii)--(iv), which link the condition to semivolatility and volatility and make it checkable through the kernel of the covariance matrix. \\
Assumption~2; ill-posed threshold models when it fails (Section~2.3)
& $\tau\in L^2$, possibly random
& Positivity on $\Delta$ in \citet{S-cetingoz2024risk}; Assumption~1 (below-mean form)
& Specialization
& Instance for $\rho_\tau$, $\tau$ possibly random; does not follow from Assumption~1; fails for a replicable benchmark. \\
Lemma~1: differentiability and gradient of portfolio semivolatility
& Assumption~1; any distribution
& \citet[Prop.~4.8, Cor.~4.9]{S-fischer2003risk}
& Specialization
& Case $p=2$ in the portfolio weights; threshold version for $\rho_\tau$. \\
Existence and uniqueness for Problem~(4) (Appendix~B of the main text)
& Assumption~1 or~2; $b_i>0$
& \citet[Theorems~1 and~2]{S-cetingoz2024risk}
& Specialization
& Self-contained proof for semivolatility and for $\rho_\tau$. \\
Prop.~3: the logarithmic program realizes the risk budgets
& Assumption~1 or~2
& \citet{S-maillard2010properties} (volatility); \citet{S-cesarone2017minimum} (CVaR); \citet{S-cetingoz2024risk}
& Specialization
& Degree one normalizes the contributions; semivariance gives the same portfolio. \\
Lemma~2: reduction of degree-one ratio models
& $\rho$ convex, degree one, positive on a cone $C$
& \citet[Props.~4--5]{S-stoyanov2007optimal}; \citet[Sec.~3]{S-cesarone2025return}
& Recalled
& Restatement with positivity required only on $C$. \\
Reformulation~(10) of the most-diversified portfolio
& Assumption~1 or~2; $\eta\le\max_i\mu_i$
& \citet[Sec.~3]{S-cesarone2025return}
& Specialization
& Below-mean, zero and benchmark versions under one lemma. \\
Reformulation~(15) of the Sortino ratio
& Assumption~1 or~2; some $\widetilde\mu_i>0$
& \citet{S-stoyanov2007optimal}
& Specialization
& Below-mean and random-threshold ratios under one lemma. \\
Lemma~3: additivity for two returns
& $X,Y\in L^2$
& \citet[Lemma~2]{S-ararat2021mad} (MAD)
& New
& Downside proportionality: stronger than MAD, weaker than volatility additivity. \\
Theorem~1: multivariate additivity
& $Y_1,\dots,Y_n\in L^2$
& \citet[Theorem~6]{S-ararat2021mad} (MAD)
& New
& Common downside factor form. \\
Prop.~4: closed-form risk-parity portfolio
& Assumption~1; Theorem~1(iv)
& \citet[Prop.~13]{S-ararat2021mad} (MAD)
& New
& Immediate corollary of Theorem~1; limiting case without downside diversification, structural relevance only. \\
Degree $\times$ threshold taxonomy (Tables~1 and~3)
& ---
& Properties classical \citep{S-ogryczak1999from,S-fischer2003risk}
& New
& Degree required by each model class, with the threshold as a second axis. \\
\bottomrule
\end{tabular}
\end{adjustbox}
\end{table}

\subsection{Numerical accuracy of the optimal portfolios}
\label{S-sec:accuracy}

Table~\ref{S-tab:accuracy} complements the exit flags reported in Section~4.2 of the main text with the residuals of the constraints and of the risk-budgeting conditions, computed from the stored weights of every rolling window of the baseline experiment.
A positive exit flag is $1$ when the first-order optimality measure falls below the solver tolerance and $2$ when the step does, so that only flag $1$ certifies a stationary point.
Flag $2$ is returned in $77$ to $84\%$ of the model-window pairs, mostly by the \texttt{fmincon} programs, whose first-order optimality measure never exceeds $3.2\times10^{-7}$ on the accepted solutions, against $1.5\times10^{-12}$ for the quadratic programs.
Across the seventeen models and the two universes, the budget constraint $\mathbf{1}^\top x=1$ holds to within $8\times10^{-16}$ and no weight is negative, as must be the case, since the accepted weights are clipped at zero and renormalized (Section~4.2 of the main text); these residuals only confirm that no invalid solution was stored.
The covariance matrix is estimated with divisor $D_L-1$ and the semideviations with divisor $D_L$, which leaves every optimal portfolio unchanged, since each model is invariant to a positive rescaling of its risk functional.
For the four risk-budgeting strategies we recompute, on the in-sample returns of each window, the relative risk contributions $\mathrm{RC}_i(x)/\rho(x)$ with respect to the risk functional of the strategy, namely volatility, below-mean semivolatility or the semideviation $\rho_\tau$ with $\tau=0$ or $\tau_t=R^I_t$, and report the largest deviation from the equal budget $1/n$, relative to $1/n$.
The deviation never exceeds $3\times10^{-6}$ for the volatility, below-mean and zero-threshold strategies and $1.6\times10^{-4}$ for the benchmark threshold, so that the solutions realize the prescribed budgets to solver precision.

The last four columns measure how far each window is from violating Assumption~2.
The margin $m^\star=\min_{x\in\Delta}\max_t(\tau_t-r_t^\top x)$, computed by the linear program described in Section~4.2 of the main text, is the smallest worst daily shortfall below the threshold that a long-only portfolio can achieve, and is positive if and only if Assumption~2 holds in the window.
The quantity $\min_{x\in\Delta}\rho_\tau(x)$, computed by quadratic programming, is the smallest in-sample daily semideviation below the threshold over the simplex.
Both are between about $12$ and $26$ times smaller for the benchmark threshold than for $\tau=0$, depending on the universe and on the measure: the benchmark-relative models are well posed in every window, but closer to degeneracy, in part because active returns are less dispersed than returns, consistently with the cautious reading of their results in the main text.

\begin{table}[htbp]
\centering
\caption{Numerical accuracy and distance from degeneracy over the rolling
windows of the baseline experiment. Column~2: largest violation of the budget
constraint over the seventeen models. Columns~3--6: largest relative deviation
$\max_{i,q}|n\,\mathrm{RC}_i/\rho-1|$ of the in-sample risk contributions of the
risk-budgeting strategies from the equal budget. Columns~7--10: smallest value
over the windows of the margin $m^\star$ and of $\min_{x\in\Delta}\rho_\tau(x)$,
in daily return units, for $\tau=0$ and $\tau_t=R^I_t$.}
\label{S-tab:accuracy}
\footnotesize
\setlength{\tabcolsep}{3.5pt}
\begin{adjustbox}{max width=\textwidth,max totalheight=0.8\textheight}
\begin{tabular}{lccccccccc}
\toprule
& & \multicolumn{4}{c}{Risk-contribution deviation} & \multicolumn{2}{c}{$\min_q m^\star$} & \multicolumn{2}{c}{$\min_q\min_\Delta\rho_\tau$} \\
\cmidrule(lr){3-6}\cmidrule(lr){7-8}\cmidrule(lr){9-10}
Universe & $|\mathbf{1}^\top x-1|$ & RP-Vol & RP-SVol$_\mu$ & RP-SVol$_0$ & RP-SVol$_{R^I}$ & $\tau=0$ & $R^I$ & $\tau=0$ & $R^I$ \\
\midrule
DowJones  & 7.8e-16 & 2.0e-6 & 3.0e-6 & 2.5e-6 & 1.6e-4 & 1.44e-2 & 5.61e-4 & 2.87e-3 & 1.78e-4 \\
Nasdaq100 & 4.4e-16 & 1.2e-6 & 2.1e-6 & 2.2e-6 & 9.8e-5 & 1.32e-2 & 7.45e-4 & 3.20e-3 & 2.58e-4 \\
\bottomrule
\end{tabular}
\end{adjustbox}
\end{table}

\setcounter{section}{0}
\renewcommand{\thesection}{S.\Alph{section}}
\renewcommand{\thesubsection}{S.\Alph{section}.\arabic{subsection}}
\section{Estimators}
\label{S-app:est}

In this appendix $T$ denotes the length of the series to which an estimator is applied.

\subsection{HAC covariance}
For a $T\times k$ matrix $Y$ of moment conditions,
\begin{equation}
\widehat V=\widehat\Gamma_0
+\sum_{j=1}^{L}\Bigl(1-\tfrac{j}{L+1}\Bigr)
\bigl(\widehat\Gamma_j+\widehat\Gamma_j^{\top}\bigr),\qquad
\widehat\Gamma_j=\tfrac1T\sum_t(y_t-\bar y)(y_{t-j}-\bar y)^{\top},
\end{equation}
with the fixed bandwidth $L=\lfloor4(T/100)^{2/9}\rfloor$, the initial lag of the automatic procedure of \citet{S-newey1994automatic}.

\subsection{Delta method for the Sharpe ratio}
\label{S-app:delta}
Let $r$ denote the daily out-of-sample return of a series and $\theta=(m_1,m_2)$ with $m_1=\mathbb{E}[r]$ and $m_2=\mathbb{E}[r^2]$.
Then
\begin{equation}
\mathrm{SR}=\frac{m_1}{\sigma},\ \sigma^2=m_2-m_1^2:\qquad
\frac{\partial \mathrm{SR}}{\partial m_1}=\frac{m_2}{\sigma^3},\qquad
\frac{\partial \mathrm{SR}}{\partial m_2}=-\frac{m_1}{2\sigma^3}.
\end{equation}
The variance of the difference between two such statistics is $\nabla^{\top}\widehat V\,\nabla/T$ over the four moment conditions involved.
These are daily derivatives; with a zero risk-free rate the annualized ratio is $\sqrt{252}\,\mathrm{SR}$, so that differences and standard errors scale by $\sqrt{252}$ and the studentized statistic and the $p$-values do not change.
The test is that of \citet{S-jobson1981performance} and \citet{S-memmel2003performance} with the HAC studentized bootstrap of \citet{S-ledoit2008robust}, who use a prewhitened quadratic spectral kernel and a block-based standard error, whereas here a Bartlett HAC matrix $\widehat V$ is recomputed at each replication; Table~\ref{S-tab:sensblock} shows that the margins are insensitive to the bandwidth.

\subsection{Circular block bootstrap}
A total of $\lceil T/b\rceil$ blocks of length $b=21$ are drawn with starting points uniform on the sample wrapped on a circle, and the concatenation is truncated to length $T$, so that every observation is equally likely to be drawn and within-block dependence is preserved \citep{S-politis1992circular}.

\renewcommand{\refname}{References for the Supplementary Material}

\end{document}